\documentclass[aps,pra,reprint,a4paper,superscriptaddress,numbers,longbibliography,showpacs,showkeys]{revtex4-1}

\usepackage[pdftex]{hyperref,color,graphicx}
\usepackage{amsfonts,amssymb,amsmath,mathtools}
\usepackage[separate-uncertainty=false]{siunitx}
\usepackage[english]{babel}
\usepackage[utf8]{inputenc}
\usepackage[T1]{fontenc}
\usepackage{xspace}

\usepackage{framed}

\hypersetup{
    pdftitle = {Fast, high-precision optical polarization synthesizer for ultracold atom experiments},
    pdfauthor = {Carsten Robens,  Stefan Brakhane, Wolfgang Alt, Dieter Meschede, Jonathan Zopes, and Andrea Alberti},
	colorlinks=true,linkcolor=blue,citecolor=blue,filecolor=blue,urlcolor=blue
}

\DeclareMathOperator{\tr}{tr}

\newif\ifusebibfile
\usebibfiletrue

\newcommand{\figref}[2]{\hyperref[#1]{\ref{#1}(#2)}}

\def\ket#1{\mathinner{|{#1}\rangle}}

\newcommand{\spinup}{\ket{{\uparrow}}}
\newcommand{\spindown}{\ket{\downarrow}}

\newcommand*\valuesep{\,}

\newcommand\footnoteref[1]{\protected@xdef\@thefnmark{\ref{#1}}\@footnotemark}

\newcommand{\SOP}{state of polarization\xspace}
\renewcommand{\SOP}{SOP\xspace}

\newcommand{\encapsulateMath}[1]{\raisebox{0pt}[0pt][0pt]{#1}}

\def\PS{polarization-synthesized\xspace}
\def\DOP{\text{DOP}}
\def\RIN{\text{RIN}}

\newcommand*\numerr[2][]{%
  \begingroup
  \sisetup{#1}%
  \ensuremath{\pm \num{#2}}%
  \endgroup
}
\newcommand\SIerr[4][]{%
  \begingroup
  \sisetup{#1}%
  \ensuremath{(\num{#2}%
    \valuesep\numerr{#3})
    \valuesep\si{#4}}%
  \endgroup
}

\DeclareSIUnit{\dBc}{\text{dBc}}
\DeclareSIUnit\deg{deg}
\DeclareSIUnit\quanta{quanta}
\DeclareSIUnit\quant{quant}
\DeclareSIUnit\gauss{G}
\DeclareSIUnit\centimeter{cm}

\addto\captionsenglish{}
\addto\captionsenglish{}

\renewcommand\textemdash{\leavevmode\unskip\kern0.8pt\rule[0.19\baselineskip]{8pt}{0.4pt}\kern1pt\ignorespaces}

\begin{document}
\textheight=24.1cm \selectlanguage{english}

\title{Fast, high-precision optical polarization synthesizer\texorpdfstring{\\}{ }for ultracold atom experiments}

\author{Carsten Robens$^{1}$, Stefan Brakhane$^{1}$, Wolfgang Alt$^{1}$, Dieter Meschede$^{1}$, Jonathan Zopes$^{1}$, and Andrea Alberti} \email{alberti@iap.uni-bonn.de}  
\affiliation{Institut für Angewandte Physik, Universität Bonn, Wegelerstr.~8, D-53115 Bonn, Germany}

\begin{abstract}
We present a novel approach
to precisely synthesize arbitrary polarization states of light with a high modulation bandwidth.
Our approach consists of superimposing two laser light fields with the same wavelength, but with opposite circular polarizations, where the phase and the amplitude of each light field are individually controlled.
We find that the polarization-synthesized beam reaches a degree of polarization of \SI{99.99}{\percent}, which is mainly limited by static spatial variations of the polarization state over the beam profile. We also find that the depolarization caused by temporal fluctuations of the polarization state is about $\num{2}$ orders of magnitude smaller.
In a recent work, Robens \emph{et al.}~[\href{http://dx.doi.org/10.1103/PhysRevLett.118.065302}{Phys.\ Rev.\ Lett.\ \textbf{118}, 065302 (2017)}] demonstrated an application of the polarization synthesizer to create two independently controllable optical lattices, which trap atoms depending on their internal spin state.
We here use ultracold atoms in \PS optical lattices to give an independent, \emph{in situ} demonstration of the  performance of the polarization synthesizer.
\end{abstract}
\maketitle

\section{Introduction}

Dynamical polarization control of light fields plays an important role in photonic applications, and it has recently been gaining importance in quantum technologies as well (see, e.g., Refs.~\cite{Mandel03,Mandel04,Karski09,Steffen12,Li:2012,Genske:2013,Steffen:2013,Zhu:2013,LeKien:2013}).
Static polarization control is much less demanding, and can be simply achieved using a few birefringent optical elements: A half- and a quarter-wave plate are already sufficient to transform a linear polarization state into any desired polarization state.

Currently existing devices for the dynamical polarization synthesis are based on voltage-controlled retarders \textemdash implemented by either  fiber squeezers or electro-optical modulators{} \textemdash and are typically specified to reach  modulation bandwidths of $\SI{100}{\kilo\hertz}$ with \SI{1}{\degree} uncertainty in the state of polarization (SOP) and \SI{99}{\percent} degree of polarization (DOP).
In general, these devices allow one to create any \SOP, but only a few of them also permit an endless, reset-free rotation of \SOP, which is achieved, e.g., by cascading multiple retarders steered  via advanced algorithms~\cite{Walker:1990,Oswald:2006}.
Polarization synthesizers of this kind are widely used in fiber-based telecommunication technologies~\cite{chen:2016}, where slow drifts of the polarization state must be actively counteracted.

The demands imposed by quantum-technological applications in terms of modulation bandwidth and precision
often go beyond the reach of existing polarization synthesizers.
Previous results of ours \cite{Karski:2011} demonstrated dynamical rotations of the linear polarization of light with a bandwidth of $\lesssim\SI{400}{\kilo\hertz}$ and a DOP at around $\SI{99.9}{\percent}$, limited by static polarization inhomogeneities across the beam profile.
While these values outperform most commercial polarization synthesizers, higher DOPs may be required~\cite{Alberti14} to suppress decoherence caused by spatial inhomogeneities and temporal fluctuations of the SOP, with the ultimate goal of achieving complex quantum manipulations of ultracold atoms comprising hundreds of quantum gates. Moreover, a single electro-optical modulator used to rotate the SOP does not permit to also control  the degree of ellipticity, and the rotation angle is limited to within a range of about $\pi$\textemdash two factors that constrain its applicability for ultracold-atom experiments.
{}

In this work, we report on a different technique for polarization synthesis, which is tailored to the requirements of ultracold atom experiments and similar quantum technologies, where polarization precision and high modulation bandwidth play an important role.
Our polarization synthesizer, instead of using an electro-optical modulator to control the SOP, directly synthesizes arbitrary SOPs by superimposing two distinct phase-stabilized laser beams with orthogonal circular polarizations.
In a recent paper \cite{Robens:2016bf}, we demonstrated an application of our polarization synthesizer to realize polarization-synthesized  optical lattices, which allow transporting atoms state dependently over arbitrarily long distances relying on a reset-free rotation of linear polarization. Thereby, we were able to demonstrate sorting of individual atoms to predefined patterns, thereby reducing the positional entropy of a randomly distributed ensemble to virtually zero.
Furthermore,  \PS{} optical lattices  have also enabled the realization of so-called ideal negative measurements for fundamental tests of quantum superimposition states~\cite{Robens:2015,Robens:2016uz}.
We expect several other applications of \PS{} optical lattices in the realm of ultracold atoms, ranging from fast atom transport~\cite{Murphy:2009,Torrontegui:2011} to testing the indistinguishability of identical particles \cite{Roos:2017}, nonequilibrium quantum thermodynamics experiments~\cite{Dorner:2013}, nonequilibrium localization experiments~\cite{Horstmann:2010}, and the quantum simulation of quantum electrodynamics~\cite{deVega:2008} and of impurity models~\mbox{\cite{Carlos:2014,Shi:2016}}.
In addition, our experimental scheme for the synthesis of light polarization may find applications in other quantum-technological areas beyond ultracold atoms.

The article is organized as follows:
In Sec.~\ref{sec:PolarizationSynthesizer} we present the experimental setup of the polarization synthesizer and its application to \PS{} optical lattices.
In Sec.~\ref{sec:PolarizationSynthesizerCharacterization} we analyze and quantify the physical mechanisms limiting the precision of the SOP, the DOP, and the modulation bandwidth of the polarization synthesizer.
Furthermore, we utilize atoms trapped in the \PS{} optical lattice to provide complementary measurements of the heating rate and transport excitations, which give an independent assessment of the performance of the polarization synthesizer.

\section{Polarization synthesizer}
\label{sec:PolarizationSynthesizer}

\subsection{Polarization synthesis}
\begin{figure}[b]
	\includegraphics*[width=\columnwidth]{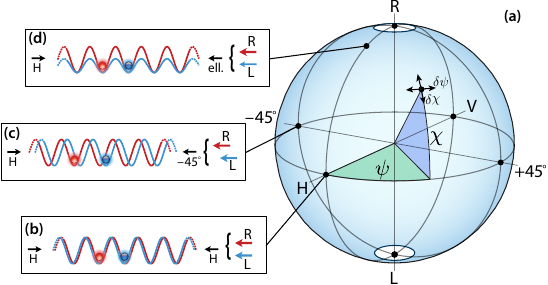}
	\caption{\label{fig:poincare} (a) Representation of the synthesized polarization state on the Poincaré sphere with rotation angle $\psi$ and ellipticity angle $\chi$.
		 The interference between the polarization-synthesized laser beam and a linearly polarized, counterpropagating laser beam gives rise to two standing waves of opposite circular polarization that are (b) spatially overlapped, (c) relatively shifted by a quarter period, and (d) of different trap depths.
Colors (red and blue) are used to denote atoms in different internal states, as well as their corresponding state-dependent optical potentials. The double-headed arrows indicate how the synthesized polarization state is affected by phase (${\delta}{}\psi$) and intensity (${\delta}{}\chi$) fluctuations of the field components in Eq.~(\ref{eqn:SynthPolEfield}).
	Circular regions close to the poles represent exclusion regions, which are not accessible by the polarization synthesizer due to the finite dynamic range of the intensity control loops.
						}
\end{figure}
The basic idea behind our polarization synthesizer is to superimpose a right (R) and a left (L) circularly polarized  laser beam, each of them with a controllable phase ($\phi_\text{R}$ or $\phi_\text{L}$) and a real-valued electric-field amplitude ($E_\text{R}$ or $E_\text{L}$), in order to produce a single laser beam with the desired polarization.
The  electric field of the resulting polarization-synthesized laser beam is given by
\begin{equation}
	\label{eqn:SynthPolEfield}
	\vec{E} = \frac{1}{\sqrt{2}}\left[E_\text{R}\begin{pmatrix} 1 \\ i \\ 0 \end{pmatrix}e^{i \phi_\text{R}} + E_\text{L} \begin{pmatrix} 1 \\ -i \\ 0 \end{pmatrix}e^{i \phi_\text{L}}\right]e^{i(kz-\omega t)}\,,
\end{equation}
where we assume a homogeneously polarized laser wavefront. Here,  $k=2\pi/\lambda$ is the wave vector, $\omega$ is the frequency of both light-field components, and the vector components of the electric field are expressed in Cartesian coordinates.
Controlling the individual phases and electric-field amplitudes allows one to synthesize any arbitrary SOP:
varying the relative phase, $\phi_\text{R}-\phi_\text{L}$,   rotates the polarization state in the real space around the laser beam's direction by an angle equal to $(\phi_\text{R}-\phi_\text{L})/2$, whereas changing the ratio between the two electric field amplitudes, $E_\text{R}$ and $E_\text{L}$,  transforms the polarization state from linear to elliptical.
For example, a horizontal linear polarization is synthesized by setting $E_\text{R}=E_\text{L}$ and $\phi_\text{R} = \phi_\text{L}$.

The polarization state of the polarization-synthesized laser beam given in Eq.~\eqref{eqn:SynthPolEfield} can be conveniently expressed as a Stokes vector $(S_0,S_1,S_2,S_3)$ \cite{Budker:2010} and visualized on the  Poincaré sphere, as shown in Fig.~\figref{fig:poincare}{a}.
The rotation ($\psi$) and ellipticity ($\chi$) angles defining the orientation of the Stokes vector can be written as a function of the control parameters of the polarization synthesizer:
\begin{align}
	\label{eqn:SphericalCoordinatesPhase}
	\psi &=\tan^{-1}(S_2/S_1)=\phi_\text{R} - \phi_\text{L}\,,\\
	\label{eqn:SphericalCoordinatesInt}
	\chi &= \sin^{-1}(S_3/S_0),\quad S_3/S_0=\epsilon=\frac{E_\text{R}^2-E_\text{L}^2}{E_\text{R}^2+E_\text{L}^2},
\end{align}
where $\epsilon$ represents the amount of ellipticity, $-1<\epsilon<1$.
Hence, a change of the relative phase rotates the Stokes vector on the Poincaré sphere in a horizontal plane, whereas an imbalance of the electric field amplitudes rotates the Stokes vector in a vertical plane.
It should be noted that most of the literature (e.g., Ref.~\cite{Collett:2005}) uses a different convention for the rotation and ellipticity angles of Stokes vectors ($\psi \rightarrow 2\,\psi$, $\chi \rightarrow 2\,\chi$).

In Ref.~\cite{Robens:2016bf}, the polarization-synthesized laser beam is made to interfere with a counterpropagating, linearly polarized beam of the same frequency $\omega$.
Thereby, two optical standing waves of R and L circular polarization are produced, forming two independent optical lattices able to trap atoms in either one of two internal states.
Three examples of \PS{} optical lattices are illustrated in the insets of Fig.~\ref{fig:poincare}, corresponding to different choices for the synthesized polarization.
Note that controlling the ratio $E_\text{R}/E_\text{L}$ and the relative phase $\phi_\text{R}-\phi_\text{L}$ suffices for the purpose of synthesizing any polarization state.
However, the control of the individual phases, as well as of the individual electric field amplitudes, enables additional operations in the case in which the polarization synthesizer is used to create a \PS{} optical lattice:
For example, varying only $\phi_\text{L}$ allows one to shift the lattice potential for only one of the two internal states [see Fig.~\figref{fig:poincare}{c}], while varying $E_\text{L}$  allows one to change the corresponding lattice depth [see Fig.~\figref{fig:poincare}{d}].

While in Eq.~(\ref{eqn:SynthPolEfield}) the electric field components are assumed to be perfectly polarized, in practice, polarization imperfections reduce the DOP to less than $\num{1}$,
\begin{equation}
	\label{eqn:DOP}
	\DOP=\frac{\sqrt{S_1^2+S_2^2+S_3^2}}{S_0}<1\,.
\end{equation}
In general, there are two basic causes of depolarization \cite{Legre:2003}: (1) a mixture of spatial modes with different polarization states and (2) a mixture of spectral (temporal) modes with different polarization states.
The first cause  yields static polarization inhomogeneities, with the SOP varying stochastically across the beam profile.
The second cause instead produces temporal fluctuations of the synthesized polarization.
Such temporal fluctuations cause an uncertainty about the SOP and, correspondingly, about the orientation of the Stokes vector on the Poincaré sphere.
In this work, static and temporal fluctuations of the SOP are considered and measured separately.

\subsection{Experimental setup}
\begin{figure}[t]
	\includegraphics*[width=\columnwidth]{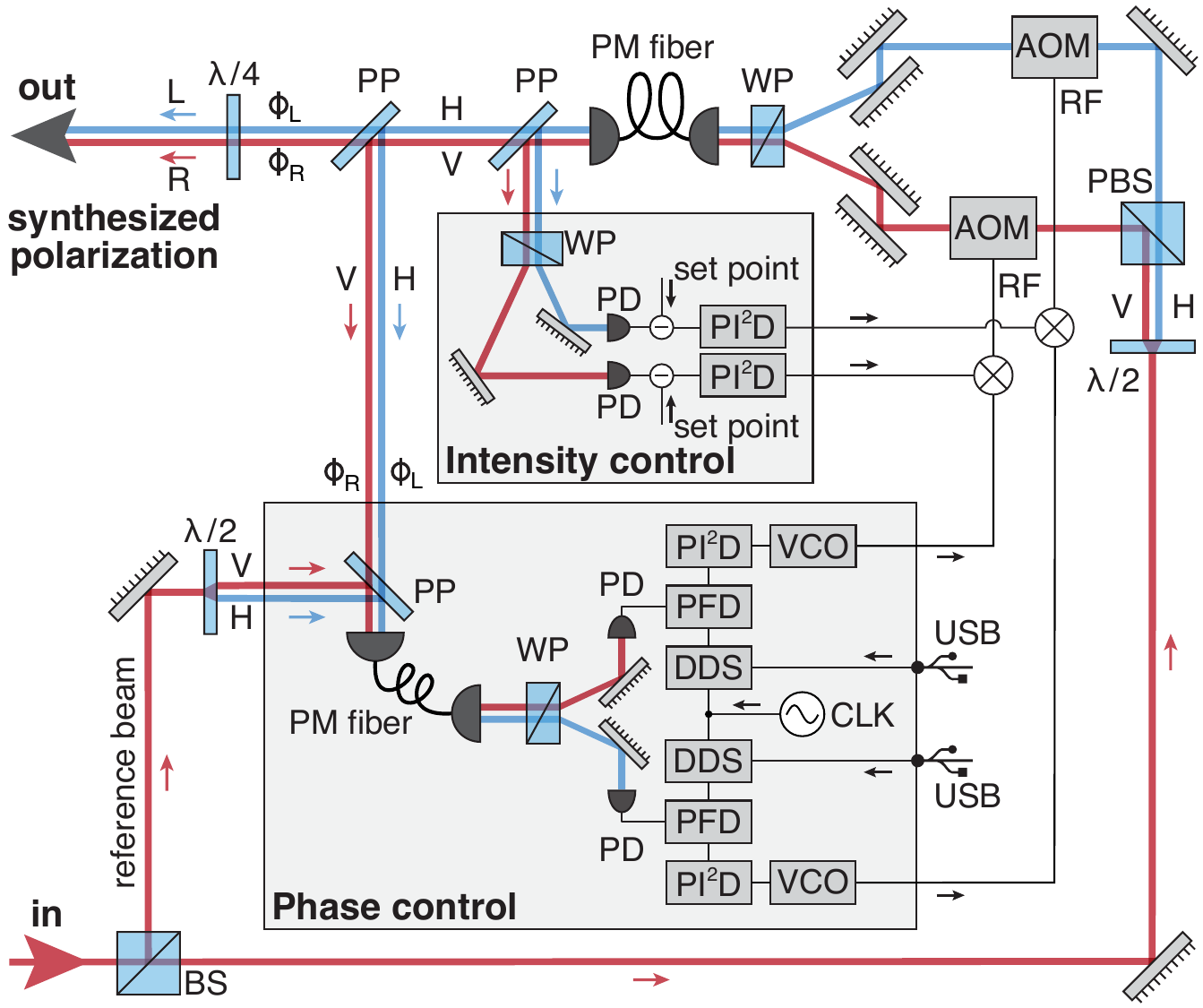}
	\caption{\label{fig:polarizationsetup} Illustration of the polarization-synthesizer setup.
	A linearly polarized laser beam enters the synthesizer at the input (in) and exits it at the output (out) with its polarization transformed into any desired SOP.
	For illustrative purposes, copropagating beams with different polarization are drawn slightly apart although they overlap in reality.
	Used abbreviations: acousto-optic modulator (AOM), beam splitter (BS), reference clock (CLK), direct digital synthesizer (DDS), horizontal linear polarization (H), polarizing beam splitter (PBS), photodiode (PD), phase frequency detector (PFD), proportional–double-integral–derivative controller (PI$^2$D), polarization maintaining (PM), pickup plate (PP), radio frequency (RF), universal serial bus (USB), vertical linear polarization (V), voltage-controlled oscillator (VCO), Wollaston prism (WP), half-wave plate ($\lambda/2$), and quarter-wave plate ($\lambda/4$).
	}
\end{figure}
Figure~\ref{fig:polarizationsetup} presents a sketch of the experimental setup for the control of the phase and amplitude of the two orthogonally polarized laser beams, which are spatially combined to synthesize the desired polarization.

The input beam from a Ti:sapphire laser (MBR 110, Coherent) is split by a beamsplitter (BS) into a reference beam required for the optical phase control and a main beam used to generate the polarization-synthesized beam.
The main beam is further divided by a polarizing beamsplitter (PBS) into two beams with vertical (V) and horizontal (H) polarization, the intensity and the phase of which are independently controlled by two separate acousto-optic modulators (AOMs).
The superimposition of the two circularly polarized light-field components in Eq.~(\ref{eqn:SynthPolEfield}) is achieved by spatially recombining both linear polarized beams with a Wollaston prism (WP) and, subsequently, by transforming the linear polarizations into circular ones using a quarter-wave plate.

We use a feedback control system in order to counteract the effect of thermal drifts, acoustic noise, air turbulence, and laser intensity noise, which cause the amplitudes ($E_\text{R}$ and $E_\text{L}$) and the phases ($\phi_\text{R}$ and $\phi_\text{L}$) of the two circularly polarized components to fluctuate.
If not properly stabilized, the phase in particular would be strongly affected by sub-wavelength mechanical vibrations of the optical components at the place where the laser beams split into separate AOMs, by the phase noise of the voltage-controlled oscillators (VCOs), and by time-varying thermal stress of the optical fiber after the Wollaston prism.
Hence, to control both phase and amplitude, we utilize for each light-field component two independent feedback control loops, indicated in Fig.~\ref{fig:polarizationsetup} by the  shaded regions, which act on the radio frequency (RF) signal sent to the AOMs.
The error signals for the control loops are obtained by diverting parts of the polarization-synthesized output beam into two beams using custom-coated  (\SI{12}{\percent} reflectivity for both polarizations) pickup plates (PP) (Altechna).

For the phase control loop, we superimpose one of two diverted beams  with the  linearly polarized reference beam mentioned at the beginning of this section.
The resulting beam is mode cleaned through a polarization-maintaining (PM) single-mode optical fiber, and the two polarization components are subsequently separated  by a Wollaston prism.
At both output ports of the Wollaston prism, beat signals are detected due to the 80-\si{\mega\hertz}-frequency difference  generated by the AOMs between the reference beam and the diverted beams.
Each beat signal is detected using an ultrafast  photodiode (G4176-03, Hamamatsu), which is AC coupled through a bias tee (ZX85-12G+, Mini-Circuits) to a low noise RF amplifier (ZFL-500HLN+, Mini-Circuits) and, subsequently, to a limiting amplifier (AD8306, Analog Devices).
The limiting amplifier strongly reduces spurious amplitude-to-phase conversions when the amplitude of the beat signal changes, whereas the low-capacitance photodiode prevents phase shifts due to changes of the capacitance induced by ambient light fluctuations.
Individually for each polarization component, the phase of the beat signal is compared to a RF reference signal (DDS) (AD9954, Analog Devices) using a digital phase-frequency discriminator (PFD) (MC100EP140, ON Semiconductor), which has an instrumental RMS phase noise measured at around $\SI{0.03}{\degree}$ over a \SI{10}{\mega\hertz} bandwidth.
The chosen DDS model allows us to store arbitrary phase ramps in its internal RAM (1024 words, 32 bits), which we use to control the phase of the reference signal in time.
The  error signal resulting from the PFD output is filtered by a 10-\si{\mega\hertz}-bandwidth proportional–double-integral–derivative (PI$^2$D) controller (D2-125, Vescent Photonics). To close the phase control loop, individually for each polarization component, the output of the analog loop filter steers through a voltage-controlled oscillator (ZX95-78+, Mini-Circuits) the frequency of the RF signal that drives the respective AOM.
The chosen VCO features a high frequency control bandwidth (5 MHz) and a low phase noise ($\SI{-140}{\dBc/\hertz}$ at $\SI{100}{\kilo\hertz}$ offset).
By closing the phase control loop, phase variations in the RF reference signal are thus imprinted onto the phase of the laser beam traversing the controlled AOM.

Three comments are in order: First, controlling the phase-locked loop (PLL) through the phase of the reference signal, instead of through the set point of the PI$^2$D, ensures that the PFD operates at around zero phase difference, where the PFD instrumental phase noise is minimum, avoids PFD nonlinearities, and, most importantly, allows us to realize reset-free phase modulations by several multiples of $2\pi$.
Second, by allowing the two polarization components to travel along a common path and by stabilizing the phase of each  component, $\phi_\text{R}$ and $\phi_\text{L}$, with respect to a common reference laser beam, we ensure precise control of the relative phase, $\psi$, of the polarization-synthesized beam [see Eq.~(\ref{eqn:SphericalCoordinatesPhase})].
Third, by employing  a common 400-\si{\mega\hertz} clock signal (CLK) as the time base for both DDS RF signal generators, we minimize the electronic contribution to the differential phase noise in the phase-control-loop setup.

For the intensity control loop, we use the second laser beam that is diverted from the polarization-synthesized beam. A Wollaston prism  spatially separates the two orthogonal polarization components. The optical power of each component is detected by a fast photodiode (PDA10A, Thorlabs) and compared to a variable set point in order to form an error signal, which is fed to an additional PI$^2$D controller (D2-125, Vescent Photonics). By means of a mixer (ZLW-6+, Mini-Circuits), the controller steers the amplitude of the RF signal used to drive the AOMs.

In order to achieve a high DOP of the synthesized polarization, static polarization inhomogeneities are strongly suppressed
by matching the transverse modes of both polarization components
through a PM single-mode optical fiber, which is situated after the Wollaston prism.
The optical fibers employed in this setup are tested to have a high linear polarization extinction ratio, $>\SI{50}{\deci\bel}$ \cite{Varnham:1984,Takada:1986,Sears:1990}.
We also find that stress-induced birefringence of the optical fiber collimators causes deviations from linear polarization unless the spurious birefringence is compensated for by an additional pair of quarter- and half-wave plates placed in front of each optical-fiber end (not shown in Fig.~\ref{fig:polarizationsetup}). This compensation ensures that the polarizations of the two electric-field components are aligned to the s and p directions of the PPs.

Moreover, in order to also ensure a high circular-polarization purity for both R and L polarization components, we use two quarter-wave plates instead of a single one at the output of the polarization synthesizer. The R and L polarization components  are  analyzed separately by blocking the other component before the Wollaston prism.
After careful adjustment of the two plates, we measure a circular-polarization purity, defined as the ratio $\mathcal{P}_{\text{R},\text{L}}=(E_\text{R}/E_\text{L})^{\pm 2}$, of $\gtrsim\num{e5}$ when only the R or L polarization component is allowed to propagate.
Under the assumption of a polarization state constant in time and homogeneous over the beam profile, one can show using the Stokes vector formalism that a finite value of the purity, $\mathcal{P}$, corresponds to synthesizing polarization states on a Poincaré-like sphere that is slightly inclined by an angle of $2\arctan(1/\sqrt{\mathcal{P}})\lesssim \SI[group-digits = false]{0.4}{\degree}$ with respect to the vertical axis.
However, the measured value of $\mathcal{P}$ is likely due to residual polarization inhomogeneities (Sec.~\ref{sec:PhaseNoisePSlattice}) caused by uncompensated for, inhomogeneous stress-induced birefringence at the output end of the PM fiber.

\subsection{Polarization-synthesized optical lattices}
	
Demonstrated in Ref.~\cite{Robens:2016bf}, polarization-synthesized optical lattices  consist of two superimposed yet independently controllable optical lattice potentials, which can trap ultracold cesium atoms depending on their internal state.
These lattice potentials are a direct application of the polarization synthesizer, which is used to create two optical standing waves with R- and L-circular polarization by making the polarization-synthesized output beam interfere with a counterpropagating, linearly polarized beam of the same frequency.
Exploiting the polarization-dependent ac polarizability of the outermost hyperfine states of cesium, namely $\spinup = \ket{\mathrm{F} = 4, \mathrm{m}_{\mathrm{F}} =4}$ and $\spindown = \ket{\mathrm{F} = 3, \mathrm{m}_{\mathrm{F}} =3}$,  at the so-called magic wavelength, $\lambda=\SI{866}{\nano\meter}$, the $\spinup$ state experiences an optical dipole potential, $U_\uparrow$, originating only from R-polarized light, while the $\spindown$ state  experiences a potential, $U_\downarrow$, generated predominantly~\cite{Belmechri:2013} by L-polarized light: 
\begin{equation}
\label{eqn:latticepotential}
	U_s(x,t) = U^{(0)}_s \cos^2\{2\pi[x-x_s(t)]/\lambda\}\,.
\end{equation}
Here, $U^{(0)}_{s}$ is the lattice depth and $x_{s}(t)$ is the longitudinal displacement of the corresponding lattice, with $s=\{\uparrow,\downarrow\}$ denoting the pseudospin orientation.
The polarization synthesizer allows us, therefore, to  control the individual positions, $x_\uparrow$ and $x_\downarrow$, of the two potentials by varying the phases $\phi_\text{R}$ and $\phi_\text{L}$ [see Fig.~\figref{fig:poincare}{c}] according to 
\begin{equation}\label{eqn:PhasePos}
	x_\uparrow(t) =  \frac{\lambda}{2} \frac{\phi_\text{R}(t)-\phi_{0}(t)}{2\pi},\quad
	x_\downarrow(t) \approx \frac{\lambda}{2} \frac{\phi_\text{L}(t)-\phi_{0}(t)}{2\pi},
\end{equation}%
where $\phi_0(t)$ is the phase of the linearly polarized counterpropagating beam, which can also be steered in time. The second equation holds only approximately due to the small contribution of R-polarized light to the lattice potential $U_\downarrow$.
Moreover, we can directly control the lattice potential depths by varying the light-field intensity $E_\text{R}$ and $E_\text{L}$, as shown in Fig.~\figref{fig:poincare}{d}.
In fact, the potential depth \encapsulateMath{$U_\uparrow^{(0)}$} is directly proportional to the light-field intensity $E_\text{R}^2$, while the potential depth \encapsulateMath{$U_\downarrow^{(0)}$} is approximately proportional to the light-field intensity $E_L^2$.
To obtain the exact expression of the position $x_\downarrow$ and potential depth \encapsulateMath{$U_\downarrow^{(0)}\!$}, the reader is referred to Ref.~\cite{Belmechri:2013}.

\section{Characterization of the polarization synthesizer}\label{sec:PolarizationSynthesizerCharacterization}
In view of future quantum applications, where particles are in fragile quantum states delocalized over many lattice sites, it is important to characterize the precision attained by the polarization synthesizer and the \PS{} optical lattice.
These characterizations are presented in detail in the following sections (\ref{sec:PhaseNoisePSlattice}, \ref{sec:PhaseNoiseHeating}, \ref{sec:PSBandwidth}, and \ref{sec:transexcitations}). We summarize the main results here.

In Sec.~\ref{sec:PhaseNoisePSlattice}, we characterize the DOP of the polarization synthesizer. For this purpose, we measure both the relative intensity noise and the relative phase noise of the two circularly polarized laser beams.
In addition, we also measure spatial polarization inhomogeneities across the profile of the polarization-synthesized beam, which also contribute to a reduction of the DOP.
The results of these measurements are summarized in Table~\ref{tab:poincare}: We find that static polarization inhomogeneities are the leading contribution degrading the polarization purity.
Our analysis reveals, furthermore, that the  noise of the relative phase, $\phi_\text{R}-\phi_\text{L}$, is particularly small, corresponding to 
a RMS uncertainty about the relative position, $\Delta x= x_\uparrow-x_\downarrow$, between the two standing waves of the \PS{} optical lattice of  $\sigma_{\Delta x}\approx\SI{1.2}{\angstrom}$.

Static polarization inhomogeneities cause state-dependent deformation of the lattice potentials, one of the main sources of inhomogeneous spin dephasing for thermal atoms or, more generally, for atoms  distributed over several motional states~\cite{Alberti14}.
By contrast, fluctuations of the synthesized polarization state due to phase and intensity noise can produce spin dephasing even for atoms cooled into the motional ground state.
However, from Ramsey interferometry \cite{Kuhr:2005}, we infer a spin-coherence time of $\SI{250}{\micro\second}$ probing thermal atoms trapped in \PS optical lattices, which is limited not by \PS{} optical lattices but by other spin-dephasing sources (stray magnetic fields, hyperfine-interaction-mediated differential light shifts; see Ref.~\cite{Kuhr:2005}).

Furthermore, fluctuations of the synthesized polarization state can also cause motional excitations.
In Sec.~\ref{sec:PhaseNoiseHeating}, we determine the heating rate from storage-time measurements, where we use a Fokker-Planck equation~\cite{Gehm:1998,Gibbons:2008,Blatt:2015} to model the loss of atoms from \PS{} optical-lattice potentials.
From our analysis of atom losses, we infer an excitation rate of about $\SI{1}{\quant/\second}$. The obtained value is consistent with the rate of excitations caused by position fluctuations of the lattice, which we estimate from the measured power spectral densities of the phase  noise. From the measured power spectral density of the intensity noise, we instead obtain that intensity noise has a negligible contribution to the heating of atoms.

Concerning the dynamical control of \PS{} optical lattices, we measure the response function of the polarization synthesizer for both the phase and intensity servo loops, obtaining a modulation bandwidth of about $\SI{800}{\kilo\hertz}$. The details of these measurements are discussed in Sec.~\ref{sec:PSBandwidth}.
Such a high bandwidth, in combination with high trapping frequencies (i.e., deep lattices), allows us to state dependently transport atoms and control their motional states on the  time scale of microseconds, which is  orders of magnitude faster than in typical quantum gas experiments.

By sideband spectroscopy, we furthermore observe that all transport operations employed in Robens \emph{et al.}~\cite{Robens:2016bf} to sort  atoms into arbitrary patterns  leave $\SI{99}{\percent}$ of the atoms in the longitudinal and transverse motional ground state (see Sec.~\ref{sec:transexcitations}).
We experimentally verify that this is the case even for nonadiabatic state-dependent transport operations lasting $\SI{20}{\micro\second}$ (corresponding to about 2 oscillation periods in the harmonic-trap approximation) per single-site shift using a bang-bang-like transport pulse in a similar manner to that employed in Ref.~\cite{Walther:2012} using trapped ions.

\subsection{Degree of polarization (DOP)}
\label{sec:PhaseNoisePSlattice}

\begin{table}[b]
	\caption{\label{tab:poincare}Physical factors limiting the degree of polarization (DOP) of the polarization-synthesized beam and their contributions to the polarization extinction ratio, $\eta$.
	Numbers in boldface are directly measured (for the experimental details, see Sec.~\ref{sec:PhaseNoisePSlattice}).
The total polarization extinction ratio reported in the table corresponds to $\DOP\approx\SI{99.99}{\percent}$, see Eq.~(\ref{eqn:etaDOP}).	%
			}
	\begin{center}
		\begin{ruledtabular}
		\begin{tabular}{lcr}
			& State of polarization\hspace{17pt} & $\eta$\hspace{7.5mm}  \\[1pt]
			\hline\\[-8pt]
			Intensity noise & $\sigma_\chi\approx \SI[math-rm = \mathbf]{0.02}{\degree}$ &$\num{4e-8}$ \\
			Phase noise & $\sigma_\psi\approx \SI[math-rm = \mathbf]{0.09}{\degree}$ & \num{6e-7} \\
			Spatial inhomogeneities& --\hspace{9.5pt} & $\num{5e-5}$\\[2pt]
			\hline\\[-8pt]
			Total&  \hspace{24.5pt}  & {$\num[math-rm = \mathbf]{5e-5}$}\\
			
		\end{tabular}
		\end{ruledtabular}
	\end{center}
													\vspace{-4mm}
\end{table}

The DOP denotes how pure the polarization state is. In real applications, polarization imperfections due to fluctuations in time and spatial inhomogeneities of the SOP reduce the DOP to less than 1, see Eq.~(\ref{eqn:DOP}).

To characterize the DOP with high accuracy, we carry out a measurement of the polarization extinction ratio~\cite{ExtinctionRatio}. We rely on the fact \cite{DOP} that the DOP is related to the minimum polarization extinction ratio,
\begin{equation}
	\label{eqn:etaDOP}
	\eta = \frac{1-\DOP}{2}\,,
\end{equation}
which can be reached after an (ideal) polarizer by adjusting the SOP in front of it using, e.g., a half- and quarter-wave plate.
Thus, we let the polarization-synthesized beam cross a polarizer (colorPol IR 1100, CODIXX), which features an extinction ratio at around $\num{e-7}$, and record with a beam profiling camera the optical power of the transmitted beam with an exposure time of $\SI{1}{\second}$.

The transmitted power, integrated over the beam profile and normalized by the power transmitted through the \SI{90}{\degree}-rotated polarizer, yields the overall polarization extinction ratio, $\eta$.
Using the dynamical control of the polarization synthesizer, we vary the rotation angle $\psi$ [see Eq.~(\ref{eqn:SphericalCoordinatesPhase})] and the ellipticity $\epsilon$ [see Eq.~(\ref{eqn:SphericalCoordinatesInt})] of the synthesized polarization to minimize $\eta$.
With this procedure, we obtain a minimum extinction ratio of $\eta \approx \num{5e-5}$, corresponding to a DOP of about \SI{99.99}{\percent}.
Note that this measurement of the DOP is sensitive not only to static spatial polarization inhomogeneities, but also to depolarization by fast temporal fluctuations of the SOP.

To obtain further insight into the factors limiting the DOP,
we analyze separately the contribution of temporal fluctuations of the control parameters, $E_{\text{R},\text{L}}$ and $\phi_{\text{R},\text{L}}$, see Eq.~(\ref{eqn:SynthPolEfield}). The details of the additional characterizations are presented below and the results are summarized in Table~\ref{tab:poincare}.

We first assume that only the intensities, $E_\text{L}^2$ and $E_\text{R}^2$, can stochastically fluctuate in time, while $\psi$ is arbitrary, yet fixed.
In the limit of small fluctuations, it can be shown using the Stokes vector formalism that the DOP is determined by the variance, $\sigma^2_\chi$, of the ellipticity angle $\chi$ and is independent of the orientation of the Stokes vector on the Poincaré sphere,
\begin{equation}
	\label{eqn:DOPchi}
	\DOP = 1- \frac{\sigma^2_\chi}{2}\,.\end{equation}
Moreover, in the same limit of small fluctuations, the previous expression in Eq.~(\ref{eqn:DOPchi}) can be related to experimentally accessible quantities,
\begin{equation}
	\label{eqn:sigmaChi}
	\sigma^2_\chi = \frac{\RIN_\text{R}^2+\RIN_\text{L}^2}{4}(1-\epsilon^2)\,,
\end{equation}
where \encapsulateMath{$\RIN_{\text{R},\text{L}} = \sigma_{E^2_{\text{R},\text{L}}}/I_{\text{R},\text{L}}$} is the relative intensity noise (RIN) of the two polarization components, which can be precisely measured. Here, \encapsulateMath{$I_{\text{R}} = \langle E^2_\text{R}\rangle$} is the average intensity (up to a constant prefactor) and \encapsulateMath{$\sigma^2_{E^2_{\text{R}}}$} is the variance of the intensity $E^2_{\text{R}}$; analogous definitions also hold for the $L$-polarized light-field component.
Note that Eq.~(\ref{eqn:sigmaChi}) is derived under the assumption that the fluctuations $\delta E_\text{R}$ and $\delta E_\text{L}$ of the electric fields $E_\text{R}$ and $E_\text{L}$ are uncorrelated, which is reasonable for noise spectral components  within the bandwidth of two independent intensity control loops.
It may be useful to also specify the two limiting cases of perfect correlations and anticorrelations, $\delta E_\text{L}=\pm \delta E_\text{R} \sqrt{I_\text{L}/I_\text{R}}$.
For perfectly correlated fluctuations, we obtain $\sigma_\chi^2=0$ (thus, $\DOP=1$) whereas for anticorrelated fluctuations we find that $\sigma_\chi^2$ amounts to twice the value given in Eq.~(\ref{eqn:sigmaChi}).

Equation~(\ref{eqn:sigmaChi}) shows that for a given amount of RIN, the DOP has the worst value (its minimum) for  linear polarization, $\epsilon = 0$, whereas the DOP is ideally 1 for a circular polarization, $\epsilon = \pm 1$, when intensity fluctuations of either the R- or L-polarized beam have no effect on the SOP.
Thus, we characterize the  depolarization of the output beam due to intensity fluctuations in the most unfavorable case of a synthesized linear polarization.
We measure the RIN separately for each of the two circularly polarized beams by integrating the intensity noise spectral density from \SI{1}{\hertz} to \SI{25}{\mega\hertz}   using a spectrum analyzer.
Both RIN measurements amount to
a similar value, $\RIN_{\text{R},\text{L}} \approx \SI{0.056}{\percent}$, resulting in $\sigma_\chi \approx \SI{0.02}{\degree}$ and, correspondingly, in a contribution to the polarization extinction ratio of about $\num{4e-8}$.

Now, we assume that only the phases, $\phi_\text{R}$ and $\phi_\text{L}$, can stochastically fluctuate in time, while the intensities are fixed.
Using the Stokes vector formalism it can be shown that, for the same limit of small fluctuations considered before, the DOP is determined by the variance, $\sigma_\psi^2$, of the rotation angle,
\begin{equation}
	\label{eqn:DOPpsi}
	\DOP =  1-\frac{1-\epsilon^2}{2}\sigma_\psi^2\,.
\end{equation}
Moreover, if we assume that the fluctuations of $\phi_\text{R}$ and $\phi_\text{L}$ are uncorrelated (which is reasonable for noise spectral components in the bandwidth of the phase control loop), we directly obtain [see Eq.~(\ref{eqn:SphericalCoordinatesPhase})] \encapsulateMath{$\sigma_\psi^2=\sigma_{\phi_\text{R}}^2 + \sigma_{\phi_\text{L}}^2$}, where \encapsulateMath{$\sigma^2_{\phi_\text{R}}$} and \encapsulateMath{$\sigma^2_{\phi_\text{L}}$} are the variances of the individual phases, respectively.
Note also that the DOP here depends on the orientation of the Stokes vector, namely, on its ellipticity, in contrast to the expression in Eq.~(\ref{eqn:DOPchi}).
The reason for this dependence becomes apparent if we consider limiting cases that are R- or L-circularly polarized, situation in which the fluctuations of the rotation angle, $\psi$, cannot cause depolarization.

\begin{figure}[t]
	{\includegraphics*[width=\columnwidth]{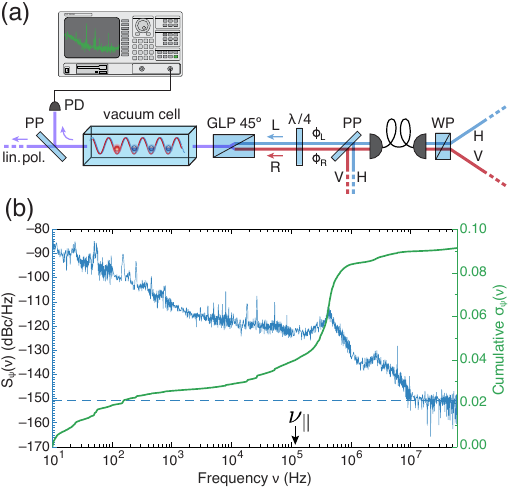}}%
	\caption{\label{fig:phasenoise}(a) Optoelectronic setup to measure the noise spectral density, $S_\psi(\nu)$, of the relative phase, $\psi = \phi_\text{R}-\phi_\text{L}$, between the R- and the L-polarized laser beams:
With respect to the setup in Fig.~\ref{fig:polarizationsetup}, a Glan-laser polarizer (GLP) oriented at \SI{45}{\degree} is present between the quarter-wave plate and the vacuum cell, which allows us
	to analyze the phase noise of the polarization-synthesized optical lattice directly \emph{in situ}.
	(b) 
	The measured phase noise spectral density is shown in units of \si{\dBc/\hertz} \cite{SingleSidebandPhaseNoise} (left axis) over more than six decades in frequency, with the electronic noise floor subtracted.
	The cumulative RMS phase noise, $\sigma_\psi(\nu)$, is also displayed (right axis).   The signal is $>\SI{20}{\decibel}$ above the electronic noise floor for $\nu<\SI{1}{\mega\hertz}$, and the noise floor lies at about $\SI{-152}{\dBc/\hertz}$ for $\nu>\SI{10}{\kilo\hertz}$; from the transimpedance of the photodiode, we estimate the photon shot noise to lie at $\SI{-150.5}{\dBc/\hertz}$ (the horizontal dashed line). The longitudinal oscillation frequency of atoms trapped in the polarization-synthesized optical lattice is indicated by~$\nu_\parallel$, where the phase noise is $S_{\psi}(\nu_\parallel)\approx \SI{-122}{\dBc/\hertz}$.}
\end{figure}

However, instead of measuring separately \encapsulateMath{$\sigma^2_{\phi_\text{R}}$} and \encapsulateMath{$\sigma^2_{\phi_\text{L}}$} to obtain $\sigma_\psi^2$, we directly measure the noise spectral density of the relative phase, $\psi$. For this purpose, we record the intensity fluctuations of a polarization-synthesized beam with $\epsilon=0$ after transmission through a Glan-laser polarizer oriented at $\SI{45}{\degree}$ with respect to the synthesized linear polarization, as illustrated in Fig.~\figref{fig:phasenoise}{a}.
Thereby, small  phase variations are linearly converted into intensity variations, which are recorded by a fast photodiode and Fourier analyzed by a spectrum analyzer; see Fig.~\figref{fig:phasenoise}{b}.
By integrating the phase noise spectral density from \SI{1}{\hertz} to $\SI{25}{\mega\hertz}$, we obtain   $\sigma_\psi\approx\SI{0.09}{\degree}$, which, according to Eqs.~(\ref{eqn:etaDOP}) and (\ref{eqn:DOPpsi}), results in a contribution to the polarization extinction ratio of about $\num{8e-7}$.

For the polarization-synthesized optical-lattice application, we use the measurement of $\sigma_\psi$ to obtain the RMS uncertainty, \encapsulateMath{$\sigma_{\Delta x} \approx \SI{1.2}{\angstrom}$}, about the relative position, $\Delta x = x_\uparrow-x_\downarrow$, see Eq.~(\ref{eqn:PhasePos}).
Importantly, \encapsulateMath{$\sigma_{\Delta x}$} is much smaller than the size of the wave packet of atoms prepared in the motional ground state, which amounts, typically, to $\gtrsim\SI{20}{\nano\meter}$ \cite{Belmechri:2013}.

By comparing the values summarized in Table~\ref{tab:poincare}, we realize that the intensity and the phase noise contribute about \SI{1}{\percent} of the total measured polarization extinction ratio. 
Thus, we deduce that the main factor limiting the DOP are static spatial polarization inhomogeneities.
The images of the beam profile acquired after the polarizer in an extinction configuration provide further confirmation of our findings since the extinction ratio  exhibits spatial variations of the same order of magnitude, around $\num{e-5}$.
We suggest that the observed spatial polarization inhomogeneities originate from stress-induced birefringence in the collimator of the fiber used to clean the transverse mode of the polarization-synthesized beam.

\subsection{Phase-noise-induced heating of atoms in a \PS{} optical lattice}
\label{sec:PhaseNoiseHeating}

Fluctuations of the optical phases $\phi_\text{R}$ and $\phi_\text{L}$ shake the trap's positions, $x_\uparrow$ and $x_\downarrow$, see Eqs.~\eqref{eqn:latticepotential} and \eqref{eqn:PhasePos}.
To estimate the rate of excitations induced by phase noise, we assume a one-dimensional (1D) harmonic confinement of atoms, which is a suitable approximation for molasses-cooled atoms trapped in a deep  optical lattice. 
We model the shaking as a perturbation to the harmonic trapping potential~\cite{Gehm:1998},
\begin{equation}
	U_s(x)= \frac{1}{2}\hspace{0.5pt}m\hspace{0.5pt}(2\pi\nu_\parallel)^2(x-x_s)^2,
\end{equation}
where $m$ is the mass of cesium atoms, $\nu_\parallel$ is the longitudinal trapping frequency, and $x_s$ is the trap position, which is a fluctuating quantity with noise spectral density \encapsulateMath{$S_{x_s}(\nu)$}.
Since the noise spectral density of the position is comparable for both spin components, without loss of generality we consider in the remainder of this section the internal state $s=\;\uparrow$.

Using Fermi's ``golden rule,'' one directly obtains the transition rate $R_{n\pm1\leftarrow n}$ for an atom occupying the motional level $n$ to be transferred to the $n\pm1$ level,
\begin{equation}
	\label{eqn:PhaseNoiseTransitionRate}
	R_{n\pm1\leftarrow n} = \frac{2\hspace{0.5pt}\pi^3 \hspace{0.5pt} m\,\nu_\parallel^3}{\hbar}\,S_{x_\uparrow}(\nu_\parallel)\left(n+\frac{1}{2}\pm\frac{1}{2}\right).
\end{equation}
Moreover, \encapsulateMath{$S_{x_\uparrow}(\nu)$} is directly related to the noise spectral density of $\Delta\phi_\text{R} = \phi_\text{R}-\phi_0$ [see Eq.~\eqref{eqn:PhasePos}],
\begin{equation}
 S_{\Delta\phi_\text{R}}(\nu) = 4k^2 S_{x_\uparrow}(\nu)\,,
\end{equation}
which can be precisely measured by means of a purely optical setup, as is described below.
Hence, the average excitation rate, \encapsulateMath{$\dot{Q}_{\Delta \phi_\text{R}}$}, of an atomic ensemble is given by
\begin{align}
     \dot{Q}_{\Delta \phi_\text{R}}&= \sum_{n} P(n,t) \, \hspace{0.5pt}   \,(R_{n+1\leftarrow n}-R_{n-1\leftarrow n})\nonumber \\
  	\label{eqn:PhaseNoiseDefinition}
    {} &=R_{1\leftarrow0}= \frac{\pi^3\hspace{0.5pt}m \hspace{1pt}\nu_\parallel^3}{2\hspace{0.5pt}\hbar\hspace{0.5pt}k^2}  \,S_{\Delta\phi_\text{R}}(\nu_\parallel)\,,
\end{align}
where  $P(n,t)$ denotes the probability that an atom of the ensemble occupies the $n$-th motional level.

\begin{figure}[t]
  \begin{center}
    \includegraphics[width=1.00\columnwidth]{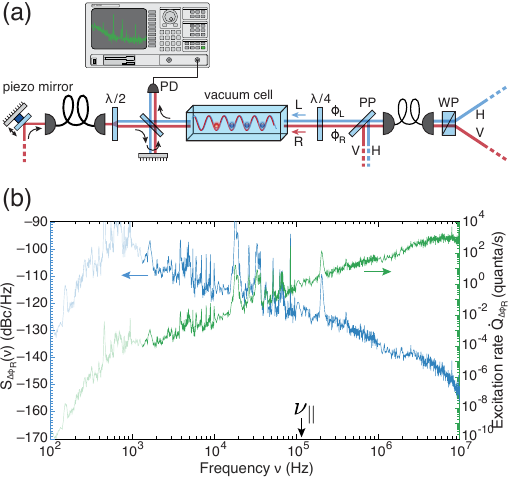}
  \end{center}
    \caption{{(a)} Optoelectronic setup to measure the phase noise values, \encapsulateMath{$S_{\Delta\phi_\text{R}}(\nu)$} and \encapsulateMath{$S_{\Delta\phi_\text{L}}(\nu)$}, of each individual optical standing wave, which are used to selectively trap atoms in the states $\ket{\uparrow}$ and $\ket{\downarrow}$, respectively.
	The two counterpropagating laser beams are made to interfere in a Michelson-like interferometer.
	The interference signal  is recorded by a fast photodiode and Fourier analyzed using a spectrum analyzer.
	A slow feedback control loop stabilizes the phase difference, $\Delta\phi_{\text{R},\text{L}} = \phi_{\text{R},\text{L}}-\phi_0$, between the two beams using a piezoelectric-actuated mirror to maintain the interference signal at the side of the fringe.
		(b) Measured phase noise spectral density, \encapsulateMath{$S_{\Delta\phi_\text{R}}(\nu)$}, and excitation rate, $\dot{Q}_{\Delta \phi_\text{R}}$, estimated using  Eq.~(\ref{eqn:PhaseNoiseDefinition}) for different trapping frequencies, $\nu$; the phase-noise-limited ground-state lifetime corresponds to $1/\dot{Q}_{\Delta \phi_\text{R}}$. The longitudinal trapping frequency of the optical lattice used in this work is indicated by $\nu_\parallel$.
	Lighter tones denote the portion of the spectrum within the bandwidth of the slow stabilization control loop.
	Differences between this spectrum and that in Fig.~\figref{fig:phasenoise}{b} (e.g., the servo bump at around $\SI{500}{\kilo\hertz}$) are likely due to a slightly different setting of the PLLs. As for Fig.~\figref{fig:phasenoise}{b}, similar considerations about the subtracted electronic noise floor apply.
\label{fig:HeatingRatesOpticalMeasurement}}
\end{figure}

Thus, in order to infer \encapsulateMath{$\dot{Q}_{\Delta \phi_\text{R}}$} from Eq.~(\ref{eqn:PhaseNoiseDefinition}), we  measure the  phase noise, \encapsulateMath{$S_{\Delta\phi_\text{R}}(\nu)$}, of one of the two optical lattice components. Note that this measurement differs from that discussed in the previous section to obtain the relative phase noise of the polarization synthesizer, \encapsulateMath{$S_{\psi}(\nu)$} [see Fig.~\figref{fig:phasenoise}{b}].
To that end, we employ the optoelectronic setup illustrated in Fig.~\figref{fig:HeatingRatesOpticalMeasurement}{a}, which consists of a Michelson interferometer where the two counterpropagating laser beams of the \PS{} optical lattice are made to interfere using a monolithic cube (W 40-4, Owis).
We use a low-bandwidth ($\lesssim\SI{1}{\kilo\hertz}$) control loop acting on a piezoelectric actuator to stabilize the interference signal at the side of the fringe, thereby ensuring that phase fluctuations are linearly converted into intensity fluctuations.

In Fig.~\figref{fig:HeatingRatesOpticalMeasurement}{b}, we show the recorded noise spectral density, as well as the excitation rate, \encapsulateMath{$\dot{Q}_{\Delta \phi_\text{R}}$}, estimated using Eq.~(\ref{eqn:PhaseNoiseDefinition}).
For a trapping frequency of $\nu_\parallel\approx\SI{117}{\kilo\hertz}$ [for its measurement, see Fig.~\figref{fig:adiabaticTransversalTransExcitations}{a}], which is typical for our quantum-walk experiments \cite{Karski09,Steffen12,Genske:2013,Robens:2015}, we obtain a phase noise $S_{\Delta\phi_\text{R}}(\nu_\parallel)\approx \SI{-122}{\dBc/\hertz}$, corresponding to an excitation rate of \encapsulateMath{$\dot{Q}_{\Delta \phi_\text{R}}\approx \SI{1}{\quanta/\second}$}.
Hence, the ground-state lifetime, $1/R_{1\leftarrow0}$, is about $\SI{1}{\second}$.

\begin{figure}[t!]
  \begin{center}
    \includegraphics[width=1.00\columnwidth]{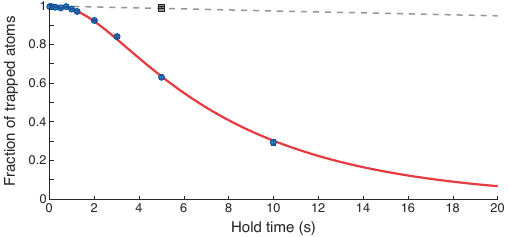}
  \end{center}
    \caption{Storage time measurement of atoms trapped in a 1D polarization-synthesized optical lattice with trap depth $U^{0}_\uparrow \approx k_\text{B}\times \SI{80}{\micro\kelvin}$.
	The circles with error bars represent the measured fraction of atoms remaining trapped after a given hold time.
	The solid line shows the result of a numerical simulation based on a Fokker-Planck equation, where the phase-noise-induced excitation rate, $\dot{Q}_{\Delta \phi_\text{R}}$, and the initial temperature of the molasses-cooled atomic ensemble, $T_0$, are fitted parameters, see Table~\ref{tab:FokkkerPlanckPDEfitResults}.
	The square point is acquired with atoms trapped in a 1D optical lattice without polarization-synthesized beam and with all AOMs supplied with the same RF signal generator.
	The dashed line shows the surviving probability of atoms, purely limited by background gas collisions, which is measured by holding atoms in a deep optical trap while they are continuously cooled by an optical molasses.
\label{fig:HeatingRatesWithFokkerPlanckHeatingFit}}
\end{figure}

To validate our estimate of the phase-noise-induced excitation rate, \encapsulateMath{$\dot{Q}_{\Delta \phi_\text{R}}$}, we carry out an independent experiment measuring the fraction of trapped atoms as a function of the  time during which the atoms are held in the \PS optical lattice without additional molasses cooling.
The measured fraction of remaining atoms is shown in Fig.~\ref{fig:HeatingRatesWithFokkerPlanckHeatingFit}, revealing a storage time (half-life) of about \SI{6.6}{\second}.
To obtain from this measurement information about \encapsulateMath{$\dot{Q}_{\Delta \phi_\text{R}}$}, we use a model of atom losses,
which considers an atom as lost once heating mechanisms have increased its energy (measured from the bottom of the trap) above the trap depth ($k_\text{B}\times\SI{80}{\micro\kelvin}$) or when the atom collides with a background gas molecule.
To account for heating mechanisms, we use a Fokker-Planck equation for a 1D harmonic trap \cite{Gehm:1998}, which describes the stochastic evolution of the energy distribution due to fluctuations of the trap position (phase noise), fluctuations of the trap depth (intensity noise), and recoil heating by the off-resonant scattering of lattice photons.
Moreover, modeling the evolution only in the longitudinal direction suffices since the rate of excitations (e.g., due to intensity fluctuations) in the directions transverse to the 1D optical lattice is significantly lower owing to the smaller transverse trapping frequency, $\nu_\perp\approx \SI{20}{\kilo\hertz}$ [for its measurement, see Fig.~\figref{fig:adiabaticTransversalTransExcitations}{b}], on which the heating rate depends \cite{Gehm:1998}.
We fit the model prediction to the experimental data (the curve in Fig.~\ref{fig:HeatingRatesWithFokkerPlanckHeatingFit}) with the initial temperature, $T_0$, of the molasses-cooled atomic ensemble and the phase-noise-induced excitation rate, \encapsulateMath{$\dot{Q}_{\Delta \phi_\text{R}}$}, being the only free parameters.
The other parameters are held fixed based on independent measurements as detailed below.
Figure~\ref{fig:HeatingRatesWithFokkerPlanckHeatingFit} shows that our model fits the measured data well.

The rate of excitation by intensity noise and the rate of losses by collisions with the background-gas molecules are too small to be derived from the fit and, thus, are provided to the model as fixed parameters based on independent measurements:
By holding atoms trapped in the optical lattice while they are continuously molasses cooled, we find that  the background-gas-limited lifetime, $\tau_\text{coll}$, of atoms in our ultra-high vacuum apparatus is about $\SI{5}{\minute}$ ($1/e$).
Moreover, from measurements of the spectrally resolved RIN (see Sec.~\ref{sec:PhaseNoisePSlattice}), \encapsulateMath{$\RIN^2_\text{R,L}(2\nu_\parallel)\approx \SI{1.9e-14}{1/\hertz}$}, we estimate \cite{Gehm:1998} that the rate constant, $\Gamma$, characterizing the exponential heating by intensity noise is about $\SI{1.3e-3}{\second^{-1}}$, corresponding to
an intensity-noise-limited ground-state lifetime, $1/R_{2\leftarrow0} = 4/\Gamma\approx \SI{50}{\minute}$, where we also take into account the RIN of the counterpropagating laser beam forming the optical lattice, which has a similar value.

Furthermore, our model does not differentiate between excitations induced by phase noise from those produced by the off-resonant scattering of lattice photons, since the excitation rates are, in both cases, independent of the atom's energy and are simply added together in the Fokker-Planck equation.
However, one can independently determine \cite{Stenholm:1986} the rate of excitations produced in the lattice direction by the recoil of the scattered photons,
\encapsulateMath{$\dot{Q}_\text{rec}=(1+1/3)\hspace{0.5pt}\gamma\hspace{0.5pt}E_\text{rec}/(2\pi\hbar\hspace{0.3pt}\nu_\parallel)$}, by knowing the recoil energy, \encapsulateMath{$E_\text{rec} =\hbar^2 k^2/(2m)$}, and the scattering rate of the lattice photons, $\gamma$.
By probing the spin relaxation induced by off-resonant scattering, we measure $\gamma \approx \SI{12.5}{\per\second}$.
Thus, we obtain \encapsulateMath{$\dot Q_\text{rec} \approx \SI{0.3}{\quanta/\second}$}, which we provide in our model as a fixed parameter.

\begin{table}
	\centering
	\caption{Results of the analysis of the storage time measurement presented in Fig.~\ref{fig:HeatingRatesWithFokkerPlanckHeatingFit}. The rate of excitations induced by phase noise, $\dot{Q}_{\Delta \phi_\text{R}}$, and the initial temperature, $T_0$, are obtained by fitting our numerical model, which relies on a Fokker-Planck equation (see the text), to the experimental data. All other quantities are independently measured and provided to the model as fixed parameters. Note that the quoted value of $1/\dot{Q}_\text{rec}$ refers to the recoil-limited lifetime in the motional ground state along the lattice direction; for the sake of completeness, we also specify the overall recoil-limited ground-state lifetime, $\SI{0.9}{\second}$, which accounts for the two transverse directions with trapping frequency $\nu_\perp$ as well.
}
	\label{tab:FokkkerPlanckPDEfitResults} 
		
		\begin{ruledtabular}
		\begin{tabular}{@{\hspace{3pt}}l@{\hspace{0pt}}r}
			\begin{tabular}{l}storage time in an optical trap of depth $U^{0}_\uparrow$\hspace*{-2cm}\end{tabular} &  \begin{tabular}{l}\SI{6.6}{\second}\end{tabular}\\[2pt]
				\hline\\[-8pt]
			\begin{tabular}{l}phase-noise-limited ground-state\\lifetime ($1/\dot{Q}_{\Delta \phi_\text{R}}$) \end{tabular}& \begin{tabular}{l}
				\SIerr{0.90}{0.05}{\second}
				\\~\end{tabular}\\[8pt]
			\hline\\[-8pt]
			\begin{tabular}{l}inverse scattering rate ($1/\gamma$)\end{tabular} &  \begin{tabular}{l}\SI{80}{\milli\second}\end{tabular}\\[2pt]
			\hline\\[-8pt]
			\begin{tabular}{l}recoil-limited ground-state lifetime ($1/\dot{Q}_\text{rec}$)\end{tabular} \hspace*{-30mm}&  \begin{tabular}{l}\SI{3.3}{\second}\end{tabular}\\[2pt]
			\hline\\[-8pt]
			\begin{tabular}{l}intensity-noise-limited ground-state\\lifetime ($4/\Gamma$)\end{tabular} &  \begin{tabular}{l} \SI{50}{\minute}\\~\end{tabular}\\[8pt]
			\hline\\[-8pt]
			\begin{tabular}{l}background-gas-limited lifetime ($\tau_\text{coll}$)\end{tabular} &   \begin{tabular}{l}\SI{5}{\minute}\end{tabular}\\[2pt]
			\hline\\[-8pt]
			\begin{tabular}{l}trap depth [$U^{0}_\uparrow={m\hspace{0.5pt}(\nu_\parallel\hspace{0.5pt}\lambda)^2}/2$]\end{tabular} &   \begin{tabular}{l}$k_\text{B}\times \SI{80}{\micro\kelvin}$\end{tabular}\\[2pt]
			\hline\\[-8pt]
			\begin{tabular}{l}atomic ensemble temperature\\after molasses cooling ($T_0$)\end{tabular} &  \begin{tabular}{l} $k_\text{B}\times\SIerr{7.8}{0.7}{\micro\kelvin}$\\~\end{tabular}\\[0pt]
		\end{tabular}
		\end{ruledtabular}
	\end{table}

The quantities determined by the fitting analysis, $T_0$ and \encapsulateMath{$\dot{Q}_{\Delta \phi_\text{R}}$}, as well as the other fixed parameters provided in the fitting model, are summarized in Table~\ref{tab:FokkkerPlanckPDEfitResults}.
The table shows that the dominant heating mechanism is phase noise.
Remarkably, the obtained value of the rate of excitations induced by phase noise, \encapsulateMath{$\dot{Q}_{\Delta \phi_\text{R}}= \SIerr{1.11}{0.06}{\quanta/\second}$}, is in good agreement with the estimate obtained from the optical measurement of phase noise; see Fig.~\figref{fig:HeatingRatesOpticalMeasurement}{b}.
The intensity noise is found to play no role in the atom losses, which are dominated by phase noise and, to a lesser extent, by the recoil heating.
Moreover, the estimated initial temperature, $T_0=\SIerr{7.8}{0.7}{\micro\kelvin}$, is in the range expected for sub-Doppler molasses cooling and agrees well with independent temperature measurements using an adiabatic-lowering technique~\cite{Alt:2003}.

To identify the primary origin of the observed phase noise, $S_{\Delta\phi_\text{R}}$, we conduct additional measurements without employing the polarization synthesizer.
For this purpose, we replace the polarization-synthesized beam with a beam of fixed linear polarization and, along with that, we supply the same RF signal to both of the AOMs employed to control the two counterpropagating optical-lattice beams.
Measurements show a remarkable suppression of the phase noise, by about 2 orders of magnitude.
Likewise, storage time measurements show a considerable increase in the fraction of trapped atoms for a given hold time (the square point in Fig.~\ref{fig:HeatingRatesWithFokkerPlanckHeatingFit}), revealing that, when the polarization synthesizer is not employed, the probability of an atom surviving in the trap is mainly determined by collisions with the background-gas molecules.
This observation gives clear evidence that the phase noise originates mostly from the employed DDS-based RF signal generators (AD9954, Analog Devices). Measurements of the electronic phase noise at the trapping frequency, $\nu_\parallel$, yield $\SI{-130}{\dBc/\hertz}$, which is only one decade lower than the measured optical phase noise, $S_{\Delta\phi_\text{R}}(\nu_\parallel)$.

However, preliminary results show that the latest generation of DDS chips (e.g., AD9915, Analog Devices) exhibits a 20-\si{\dB}-lower electronic phase noise at the trapping frequency, $\nu_\parallel$.
Employing these chips to steer the polarization synthesizer holds the promise of achieving a corresponding reduction (extension) of the heating rate (phase-noise-limited ground-state lifetime) of the trapped atoms.

To confirm that phase noise mainly originates from the DDS-based RF signal generators and not from the phase control loop itself, we repeat the measurement of the lattice phase noise, $S_{\Delta\phi_\text{R}}$, without employing the polarization synthesizer, yet using independent DDSs to drive the AOMs. In this case, we find that the amount of phase noise is comparable to that measured with the polarization-synthesized beam.

\begin{figure}[t]
	  \begin{center}
	    \includegraphics[width=1.00\columnwidth]{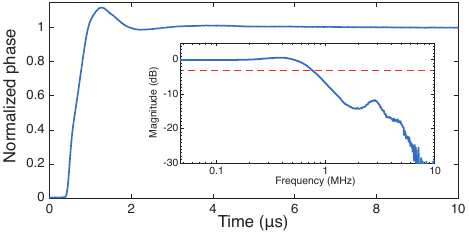}
	  \end{center}
	    \caption{Normalized closed-loop small-signal step response function of the optical phase control loop. Time $t=0$ defines the moment in which the set point phase is suddenly stepped for the R-polarized light-field component.
	    (Inset) The magnitude Bode plot of the corresponding frequency response function.
		The dashed line indicates a 3-$\si{\dB}$ attenuation.
	    } 
	    \label{fig:PhaseLockBandwidthStepResponse}
\end{figure}

\subsection{Modulation bandwidth of the polarization synthesizer}
\label{sec:PSBandwidth}

We determine the modulation bandwidth of the two intensity and two phase control loops (see Fig.~\ref{fig:polarizationsetup}) by recording their  response to a step change of the corresponding set points.
To that end, for the phase control loop, we first synthesize a linear polarization state (i.e., $\chi=0$) and then suddenly step the phase of one of the  RF signal by a small amount, say \SI{10}{\degree}.
The phase control loop reacts by rotating (in real space) the linear polarization by an angle  of $\SI{10}{\degree}/2$.
We record the dynamics of the rotation by measuring the intensity of the polarization-synthesized beam behind a $\SI{45}{\degree}$-oriented polarizer, which serves as a phase-to-intensity converter.
Figure~\ref{fig:PhaseLockBandwidthStepResponse} shows an exemplary step response function of the optical phase control loop for the R-polarized light-field component. 
From the derivative of the step response function displayed in Fig.~\ref{fig:PhaseLockBandwidthStepResponse}, we further obtain the impulse response function, whose Fourier transform, in turn, yields the frequency response function of the control loop (see inset).
From this measurement, we obtain a  modulation bandwidth (3-\si{\dB} criterion) of about $\SI{800}{\kilo\hertz}$ (the dashed red line).

For the intensity control loop, in a like manner, we record the step response function by suddenly stepping the set point intensity.
All intensity and phase control loops of the polarization-synthesizer achieve a comparable modulation bandwidth, which is primarily limited by the dead time in the response of the AOMs, which is of the order of $\SI{300}{\nano\second}$.

\subsection{Motional excitations induced by the state-dependent transport of atoms}
\label{sec:transexcitations}

\begin{figure*}[t]
  \begin{center}
    \includegraphics[width=1.00\textwidth]{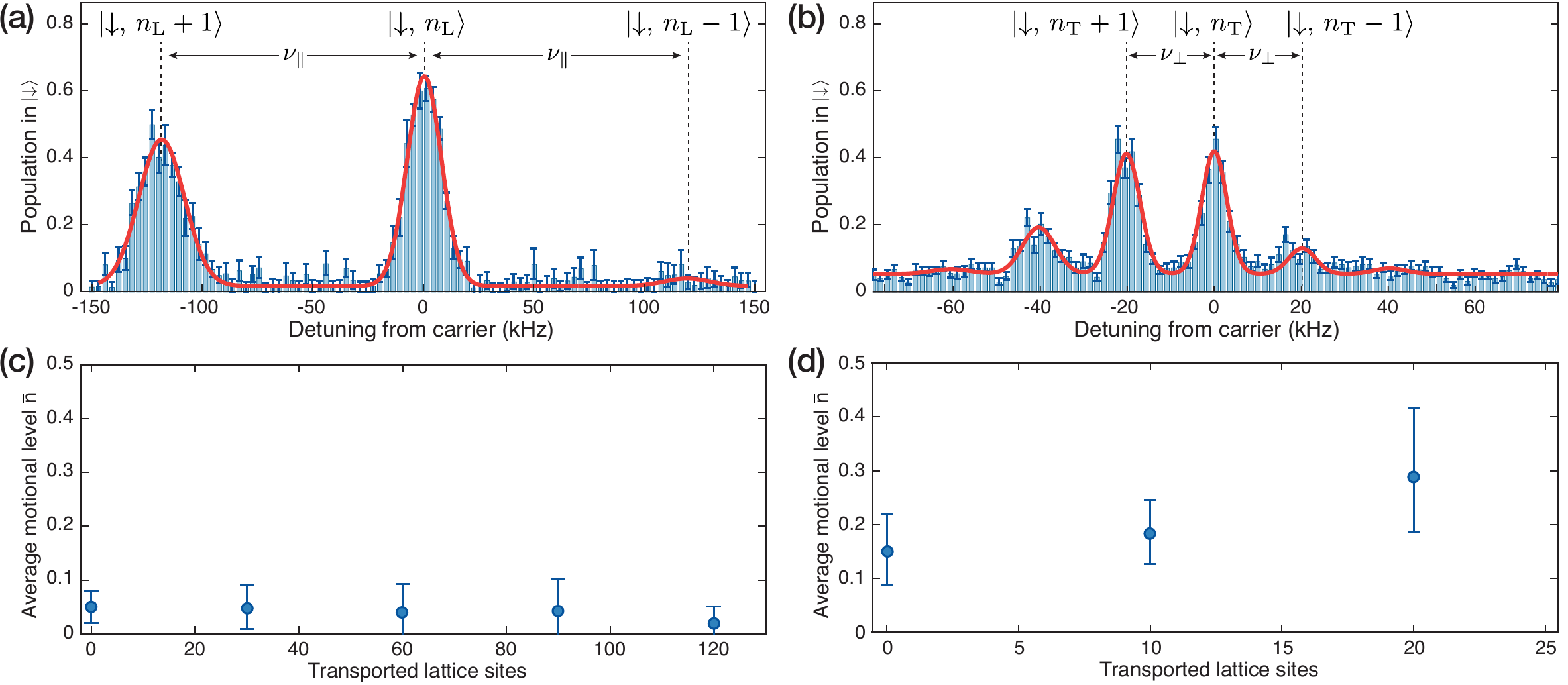}
  \end{center}
  	\vspace{-5mm}
    \caption{Typical resolved-sideband spectra of motional excitations using (a) microwave transitions for the longitudinal direction  and (b) Raman transitions for the transverse directions.
	Average motional level, $\bar{n}$, as a function of the transport distance, probing the directions (c) longitudinal and (d) transverse to the one-dimensional optical lattice.
 	} 
    \label{fig:adiabaticTransversalTransExcitations}
\end{figure*}

For applications of \PS optical lattices in which atoms are  transported (see, e.g., Refs.~\cite{Mandel03,Mandel04,Karski09,Genske:2013,Steffen12,Robens:2015,Robens:2016bf}), it is important that the transport operations do not excite atoms that are initially prepared in the motional ground state.
To transport atoms quickly (i.e., on  the time scale of $1/\nu_\parallel$), yet without creating any motional excitation, tailored transport ramps are necessary.
To that end, optimal control theory \cite{Murphy:2009} and shortcuts to adiabaticity \cite{Torrontegui:2011} provide robust solutions on how to shape both phases, $\phi_{\text{R},\text{L}}$, and intensities, $I_{\text{R},\text{L}}$.
The implementation of optimal transport will be the subject of future experimental work.

Here, we characterize motional excitations following an adiabatic transport operation.
For this purpose, we first cool the atoms into, or close to, their motional ground state by resolved-sideband cooling using microwave transitions \cite{Forster:2009,Li:2012,Belmechri:2013} (for the direction longitudinal to the lattice) and Raman transitions \cite{Han:2000,Kaufman:2012,Thompson:2013} (for the directions transverse to the lattice). Sideband cooling also initializes all atoms in state $\spinup$.
Subsequently, we translate the $U_\uparrow(x,t)$ optical-lattice potential by an integer number of lattice sites using a 1-\si{\milli\second}-long smooth ramp of the phase, $\phi_\text{R}$ [see Eqs.~(\ref{eqn:latticepotential}) and (\ref{eqn:PhasePos})]. The displacement of the lattice leads to an adiabatic acceleration and deceleration of atoms in $\spinup$, which are thereby displaced by the desired number of lattice sites.
We use motional sideband spectroscopy at the end of the transport operation \cite{Leibfried:2003,Belmechri:2013} to measure the probability of creating an excitation. 
A typical sideband spectrum is shown in Fig.~\figref{fig:adiabaticTransversalTransExcitations}{a} for the longitudinal direction, and in Fig.~\figref{fig:adiabaticTransversalTransExcitations}{b} for the transverse direction.
The three central peaks of these spectra correspond to the heating (blue) sideband transition $\ket{\uparrow,\,n}\rightarrow\ket{\downarrow,\,n+1}$, to the carrier transition $\ket{\uparrow,\,n}\rightarrow\ket{\downarrow,\,n}$, and to the cooling (red) sideband transition $\ket{\uparrow,\,n}\rightarrow\ket{\downarrow,\,n-1}$, where $n$ denotes the motional quantum number.

As a figure of merit to estimate the number of motional excitations, we use the ratio of the height of the cooling sideband to that of the heating sideband, $r$.
Assuming that the motional states are Boltzmann distributed \cite{BoltzmannDist}, this ratio is directly related to the average number of motional excitations \cite{Leibfried:2003},
\begin{equation}
	\label{eqn:meanVibrationalLevel}
	\bar{n} = \frac{r}{1-r}\,.
\end{equation}
The resulting mean motional level, $\bar{n}$, is displayed for increasing transport distances in Fig.~\figref{fig:adiabaticTransversalTransExcitations}{c} for the longitudinal direction and in Fig.~\figref{fig:adiabaticTransversalTransExcitations}{d} for the transverse direction.
We observe virtually no excitation caused by the adiabatic transport operations along the longitudinal direction and only small excitations in the transverse directions ($\lesssim\num{5e-3}$ quanta per  lattice site).

Additional measurements show excitation-free transport also for nonadiabatic transport operations using bang-bang transport pulses~\cite{Walther:2012} , which last about $\SI{20}{\micro\second}$ (i.e., two oscillation periods) per lattice site~\cite{Robens:2015}.

\section{Conclusions and outlook}

In this work, we have presented an experimental setup for the precision synthesis of arbitrary polarization states, demonstrating the capability to modulate the SOP with a bandwidth of \SI{800}{\kilo\hertz}.
Residual temporal fluctuations of the SOP, which are not suppressed by the phase and intensity control loop, limit the DOP to \SI{99.9999(1)}{\percent}.
We also find that the measured DOP of \SI{99.99}{\percent} is mainly limited by spatial inhomogeneities of the polarization state across the beam profile.
In the future, suppressing polarization inhomogeneities by, e.g., reducing stress-induced birefringence in the fiber collimator holds the promise of synthesizing polarization states with DOPs in the six 9 figures. 

Furthermore, we have shown the application of our polarization synthesizer to form state-dependent \PS{} optical lattices. Our implementation of state-dependent transport based on the polarization synthesizer overcomes the shortcomings of former realizations~\cite{Mandel03,Mandel04,Karski:2011,Steffen12}, which employed electro-optical modulators to control the SOP:
 The individual control of the two optical lattices with opposite circular polarizations not only permits fully independent shift operations of both atomic spin components, but also enables an unprecedented control of the individual lattice depths in state-dependent optical lattices. 
Such a control enables the application of optical control methods \cite{Murphy:2009} and shortcuts to adiabaticity \cite{Torrontegui:2011} to dramatically speed up transport operations.

Utilizing atoms as a measurement probe, we have provided an independent \emph{in situ} characterization of the polarization synthesizer demonstrating a remarkably low heating rate at the level of \SI{1}{\quanta/\second}, primarily limited by the phase noise of the DDS RF signal generators, and vanishing motional excitations in adiabatic atom-transport applications.

While the polarization synthesizer has been developed, in the the first place, for precise atom transport, it may find applications in other quantum technologies\textemdash or even in fiber-based telecommunication networks\textemdash to manipulate with high bandwidth and precision the polarization state of a laser beam comprising one, or a few, wavelength components.

\begin{acknowledgments}We thank A.\ Hambitzer for contributing to the experimental apparatus; S.\ Hild, A.\ Steffen, G.\ Ramola,  and J.\ M.\ Raimond for the insightful discussions; M.\ Prevedelli for suggesting to us the phase-frequency discriminator; and the anonymous referee for the valuable suggestions.
We acknowledge financial support from North Rhine-Westphalia through the Nachwuchsforschergruppe ``Quantenkontrolle auf der Nanoskala,'' the ERC grant DQSIM (project ID 291401), the EU SIQS project (project ID 600645), and the Deutsche Forschungsgemeinschaft SFB/TR 185 OSCAR.
C.R.\ acknowledges support from the Studienstiftung des deutschen Volkes, and C.R., S.B., and J.Z. from the Bonn-Cologne Graduate School.
\end{acknowledgments}

\textheight=1.08\textheight
\bibliographystyle{apsrev4-1}
\bibliography{references}

\begin{thebibliography}{48}%
\makeatletter
\providecommand \@ifxundefined [1]{%
 \@ifx{#1\undefined}
}%
\providecommand \@ifnum [1]{%
 \ifnum #1\expandafter \@firstoftwo
 \else \expandafter \@secondoftwo
 \fi
}%
\providecommand \@ifx [1]{%
 \ifx #1\expandafter \@firstoftwo
 \else \expandafter \@secondoftwo
 \fi
}%
\providecommand \natexlab [1]{#1}%
\providecommand \enquote  [1]{``#1''}%
\providecommand \bibnamefont  [1]{#1}%
\providecommand \bibfnamefont [1]{#1}%
\providecommand \citenamefont [1]{#1}%
\providecommand \href@noop [0]{\@secondoftwo}%
\providecommand \href [0]{\begingroup \@sanitize@url \@href}%
\providecommand \@href[1]{\@@startlink{#1}\@@href}%
\providecommand \@@href[1]{\endgroup#1\@@endlink}%
\providecommand \@sanitize@url [0]{\catcode `\\12\catcode `\$12\catcode
  `\&12\catcode `\#12\catcode `\^12\catcode `\_12\catcode `\%12\relax}%
\providecommand \@@startlink[1]{}%
\providecommand \@@endlink[0]{}%
\providecommand \url  [0]{\begingroup\@sanitize@url \@url }%
\providecommand \@url [1]{\endgroup\@href {#1}{\urlprefix }}%
\providecommand \urlprefix  [0]{URL }%
\providecommand \Eprint [0]{\href }%
\providecommand \doibase [0]{http://dx.doi.org/}%
\providecommand \selectlanguage [0]{\@gobble}%
\providecommand \bibinfo  [0]{\@secondoftwo}%
\providecommand \bibfield  [0]{\@secondoftwo}%
\providecommand \translation [1]{[#1]}%
\providecommand \BibitemOpen [0]{}%
\providecommand \bibitemStop [0]{}%
\providecommand \bibitemNoStop [0]{.\EOS\space}%
\providecommand \EOS [0]{\spacefactor3000\relax}%
\providecommand \BibitemShut  [1]{\csname bibitem#1\endcsname}%
\let\auto@bib@innerbib\@empty
\bibitem [{\citenamefont {Mandel}\ \emph
  {et~al.}(2003{\natexlab{a}})\citenamefont {Mandel}, \citenamefont {Greiner},
  \citenamefont {Widera}, \citenamefont {Rom}, \citenamefont {H{\"a}nsch},\
  and\ \citenamefont {Bloch}}]{Mandel03}%
  \BibitemOpen
  \bibfield  {author} {\bibinfo {author} {\bibfnamefont {O.}~\bibnamefont
  {Mandel}}, \bibinfo {author} {\bibfnamefont {M.}~\bibnamefont {Greiner}},
  \bibinfo {author} {\bibfnamefont {A.}~\bibnamefont {Widera}}, \bibinfo
  {author} {\bibfnamefont {T.}~\bibnamefont {Rom}}, \bibinfo {author}
  {\bibfnamefont {T.~W.}\ \bibnamefont {H{\"a}nsch}}, \ and\ \bibinfo {author}
  {\bibfnamefont {I.}~\bibnamefont {Bloch}},\ }\bibfield  {title} {\enquote
  {\bibinfo {title} {Coherent transport of neutral atoms in spin-dependent
  optical lattice potentials},}\ }\href
  {http://dx.doi.org/10.1103/PhysRevLett.91.010407} {\bibfield  {journal}
  {\bibinfo  {journal} {Phys. Rev. Lett.}\ }\textbf {\bibinfo {volume} {91}},\
  \bibinfo {pages} {010407} (\bibinfo {year} {2003}{\natexlab{a}})}\BibitemShut
  {NoStop}%
\bibitem [{\citenamefont {Mandel}\ \emph
  {et~al.}(2003{\natexlab{b}})\citenamefont {Mandel}, \citenamefont {Greiner},
  \citenamefont {Widera}, \citenamefont {Rom}, \citenamefont {H\"ansch},\ and\
  \citenamefont {Bloch}}]{Mandel04}%
  \BibitemOpen
  \bibfield  {author} {\bibinfo {author} {\bibfnamefont {O.}~\bibnamefont
  {Mandel}}, \bibinfo {author} {\bibfnamefont {M.}~\bibnamefont {Greiner}},
  \bibinfo {author} {\bibfnamefont {A.}~\bibnamefont {Widera}}, \bibinfo
  {author} {\bibfnamefont {T.}~\bibnamefont {Rom}}, \bibinfo {author}
  {\bibfnamefont {T.}~\bibnamefont {H\"ansch}}, \ and\ \bibinfo {author}
  {\bibfnamefont {I.}~\bibnamefont {Bloch}},\ }\bibfield  {title} {\enquote
  {\bibinfo {title} {Controlled collisions for multi-particle entanglement of
  optically trapped atoms},}\ }\href {http://dx.doi.org/10.1038/nature02008}
  {\bibfield  {journal} {\bibinfo  {journal} {Nature}\ }\textbf {\bibinfo
  {volume} {425}},\ \bibinfo {pages} {937} (\bibinfo {year}
  {2003}{\natexlab{b}})}\BibitemShut {NoStop}%
\bibitem [{\citenamefont {Karski}\ \emph {et~al.}(2009)\citenamefont {Karski},
  \citenamefont {F{\"o}rster}, \citenamefont {Choi}, \citenamefont {Steffen},
  \citenamefont {Alt}, \citenamefont {Meschede},\ and\ \citenamefont
  {Widera}}]{Karski09}%
  \BibitemOpen
  \bibfield  {author} {\bibinfo {author} {\bibfnamefont {M.}~\bibnamefont
  {Karski}}, \bibinfo {author} {\bibfnamefont {L.}~\bibnamefont {F{\"o}rster}},
  \bibinfo {author} {\bibfnamefont {J.}~\bibnamefont {Choi}}, \bibinfo {author}
  {\bibfnamefont {A.}~\bibnamefont {Steffen}}, \bibinfo {author} {\bibfnamefont
  {W.}~\bibnamefont {Alt}}, \bibinfo {author} {\bibfnamefont {D.}~\bibnamefont
  {Meschede}}, \ and\ \bibinfo {author} {\bibfnamefont {A.}~\bibnamefont
  {Widera}},\ }\bibfield  {title} {\enquote {\bibinfo {title} {{Quantum Walk in
  Position Space with Single Optically Trapped Atoms}},}\ }\href
  {http://dx.doi.org/10.1126/science.1174436} {\bibfield  {journal} {\bibinfo
  {journal} {Science}\ }\textbf {\bibinfo {volume} {325}},\ \bibinfo {pages}
  {174} (\bibinfo {year} {2009})}\BibitemShut {NoStop}%
\bibitem [{\citenamefont {Steffen}\ \emph {et~al.}(2012)\citenamefont
  {Steffen}, \citenamefont {Alberti}, \citenamefont {Alt}, \citenamefont
  {Belmechri}, \citenamefont {Hild}, \citenamefont {Karski}, \citenamefont
  {Widera},\ and\ \citenamefont {Meschede}}]{Steffen12}%
  \BibitemOpen
  \bibfield  {author} {\bibinfo {author} {\bibfnamefont {A.}~\bibnamefont
  {Steffen}}, \bibinfo {author} {\bibfnamefont {A.}~\bibnamefont {Alberti}},
  \bibinfo {author} {\bibfnamefont {W.}~\bibnamefont {Alt}}, \bibinfo {author}
  {\bibfnamefont {N.}~\bibnamefont {Belmechri}}, \bibinfo {author}
  {\bibfnamefont {S.}~\bibnamefont {Hild}}, \bibinfo {author} {\bibfnamefont
  {M.}~\bibnamefont {Karski}}, \bibinfo {author} {\bibfnamefont
  {A.}~\bibnamefont {Widera}}, \ and\ \bibinfo {author} {\bibfnamefont
  {D.}~\bibnamefont {Meschede}},\ }\bibfield  {title} {\enquote {\bibinfo
  {title} {{A digital atom interferometer with single particle control on a
  discretized spacetime geometry}},}\ }\href
  {http://dx.doi.org/10.1073/pnas.1204285109} {\bibfield  {journal} {\bibinfo
  {journal} {Proc. Natl. Acad. Sci. U.S.A.}\ }\textbf {\bibinfo {volume}
  {109}},\ \bibinfo {pages} {9770} (\bibinfo {year} {2012})}\BibitemShut
  {NoStop}%
\bibitem [{\citenamefont {Li}\ \emph {et~al.}(2012)\citenamefont {Li},
  \citenamefont {Corcovilos}, \citenamefont {Wang},\ and\ \citenamefont
  {Weiss}}]{Li:2012}%
  \BibitemOpen
  \bibfield  {author} {\bibinfo {author} {\bibfnamefont {X.}~\bibnamefont
  {Li}}, \bibinfo {author} {\bibfnamefont {T.~A.}\ \bibnamefont {Corcovilos}},
  \bibinfo {author} {\bibfnamefont {Y.}~\bibnamefont {Wang}}, \ and\ \bibinfo
  {author} {\bibfnamefont {D.~S.}\ \bibnamefont {Weiss}},\ }\bibfield  {title}
  {\enquote {\bibinfo {title} {{3D Projection Sideband Cooling}},}\ }\href
  {http://dx.doi.org/10.1103/PhysRevLett.108.103001} {\bibfield  {journal}
  {\bibinfo  {journal} {Phys. Rev. Lett.}\ }\textbf {\bibinfo {volume} {108}},\
  \bibinfo {pages} {103001} (\bibinfo {year} {2012})}\BibitemShut {NoStop}%
\bibitem [{\citenamefont {Genske}\ \emph {et~al.}(2013)\citenamefont {Genske},
  \citenamefont {Alt}, \citenamefont {Steffen}, \citenamefont {Werner},
  \citenamefont {Werner}, \citenamefont {Meschede},\ and\ \citenamefont
  {Alberti}}]{Genske:2013}%
  \BibitemOpen
  \bibfield  {author} {\bibinfo {author} {\bibfnamefont {M.}~\bibnamefont
  {Genske}}, \bibinfo {author} {\bibfnamefont {W.}~\bibnamefont {Alt}},
  \bibinfo {author} {\bibfnamefont {A.}~\bibnamefont {Steffen}}, \bibinfo
  {author} {\bibfnamefont {A.~H.}\ \bibnamefont {Werner}}, \bibinfo {author}
  {\bibfnamefont {R.~F.}\ \bibnamefont {Werner}}, \bibinfo {author}
  {\bibfnamefont {D.}~\bibnamefont {Meschede}}, \ and\ \bibinfo {author}
  {\bibfnamefont {A.}~\bibnamefont {Alberti}},\ }\bibfield  {title} {\enquote
  {\bibinfo {title} {{Electric Quantum Walks with Individual Atoms}},}\ }\href
  {http://dx.doi.org/10.1103/PhysRevLett.110.190601} {\bibfield  {journal}
  {\bibinfo  {journal} {Phys. Rev. Lett.}\ }\textbf {\bibinfo {volume} {110}},\
  \bibinfo {pages} {190601} (\bibinfo {year} {2013})}\BibitemShut {NoStop}%
\bibitem [{\citenamefont {Steffen}\ \emph {et~al.}(2013)\citenamefont
  {Steffen}, \citenamefont {Alt}, \citenamefont {Genske}, \citenamefont
  {Meschede}, \citenamefont {Robens},\ and\ \citenamefont
  {Alberti}}]{Steffen:2013}%
  \BibitemOpen
  \bibfield  {author} {\bibinfo {author} {\bibfnamefont {A.}~\bibnamefont
  {Steffen}}, \bibinfo {author} {\bibfnamefont {W.}~\bibnamefont {Alt}},
  \bibinfo {author} {\bibfnamefont {M.}~\bibnamefont {Genske}}, \bibinfo
  {author} {\bibfnamefont {D.}~\bibnamefont {Meschede}}, \bibinfo {author}
  {\bibfnamefont {C.}~\bibnamefont {Robens}}, \ and\ \bibinfo {author}
  {\bibfnamefont {A.}~\bibnamefont {Alberti}},\ }\bibfield  {title} {\enquote
  {\bibinfo {title} {{Note: In situ measurement of vacuum window birefringence
  by atomic spectroscopy.}}}\ }\href {http://dx.doi.org/10.1063/1.4847075}
  {\bibfield  {journal} {\bibinfo  {journal} {Rev. Sci. Instrum.}\ }\textbf
  {\bibinfo {volume} {84}},\ \bibinfo {pages} {126103} (\bibinfo {year}
  {2013})}\BibitemShut {NoStop}%
\bibitem [{\citenamefont {Zhu}\ \emph {et~al.}(2013)\citenamefont {Zhu},
  \citenamefont {Solmeyer}, \citenamefont {Tang},\ and\ \citenamefont
  {Weiss}}]{Zhu:2013}%
  \BibitemOpen
  \bibfield  {author} {\bibinfo {author} {\bibfnamefont {K.}~\bibnamefont
  {Zhu}}, \bibinfo {author} {\bibfnamefont {N.}~\bibnamefont {Solmeyer}},
  \bibinfo {author} {\bibfnamefont {C.}~\bibnamefont {Tang}}, \ and\ \bibinfo
  {author} {\bibfnamefont {D.~S.}\ \bibnamefont {Weiss}},\ }\bibfield  {title}
  {\enquote {\bibinfo {title} {{Absolute Polarization Measurement Using a
  Vector Light Shift}},}\ }\href
  {http://dx.doi.org/10.1103/PhysRevLett.111.243006} {\bibfield  {journal}
  {\bibinfo  {journal} {Phys. Rev. Lett.}\ }\textbf {\bibinfo {volume} {111}},\
  \bibinfo {pages} {243006} (\bibinfo {year} {2013})}\BibitemShut {NoStop}%
\bibitem [{\citenamefont {Le~Kien}\ \emph {et~al.}(2013)\citenamefont
  {Le~Kien}, \citenamefont {Schneeweiss},\ and\ \citenamefont
  {Rauschenbeutel}}]{LeKien:2013}%
  \BibitemOpen
  \bibfield  {author} {\bibinfo {author} {\bibfnamefont {F.}~\bibnamefont
  {Le~Kien}}, \bibinfo {author} {\bibfnamefont {P.}~\bibnamefont
  {Schneeweiss}}, \ and\ \bibinfo {author} {\bibfnamefont {A.}~\bibnamefont
  {Rauschenbeutel}},\ }\bibfield  {title} {\enquote {\bibinfo {title}
  {{Dynamical polarizability of atoms in arbitrary light fields: general theory
  and application to cesium}},}\ }\href
  {http://dx.doi.org/10.1140/epjd/e2013-30729-x} {\bibfield  {journal}
  {\bibinfo  {journal} {Eur. Phys. J. D}\ }\textbf {\bibinfo {volume} {67}},\
  \bibinfo {pages} {1023} (\bibinfo {year} {2013})}\BibitemShut {NoStop}%
\bibitem [{\citenamefont {Walker}\ and\ \citenamefont
  {Walker}(1990)}]{Walker:1990}%
  \BibitemOpen
  \bibfield  {author} {\bibinfo {author} {\bibfnamefont {N.~G.}\ \bibnamefont
  {Walker}}\ and\ \bibinfo {author} {\bibfnamefont {G.~R.}\ \bibnamefont
  {Walker}},\ }\bibfield  {title} {\enquote {\bibinfo {title} {{Polarization
  control for coherent communications}},}\ }\href
  {http://dx.doi.org/10.1109/50.50740} {\bibfield  {journal} {\bibinfo
  {journal} {J. Lightwave Technol.}\ }\textbf {\bibinfo {volume} {8}},\
  \bibinfo {pages} {438} (\bibinfo {year} {1990})}\BibitemShut {NoStop}%
\bibitem [{\citenamefont {Oswald}\ and\ \citenamefont
  {Madsen}(2006)}]{Oswald:2006}%
  \BibitemOpen
  \bibfield  {author} {\bibinfo {author} {\bibfnamefont {P.}~\bibnamefont
  {Oswald}}\ and\ \bibinfo {author} {\bibfnamefont {C.~K.}\ \bibnamefont
  {Madsen}},\ }\bibfield  {title} {\enquote {\bibinfo {title} {{Deterministic
  analysis of endless tuning of polarization controllers}},}\ }\href
  {http://dx.doi.org/10.1109/JLT.2006.872688} {\bibfield  {journal} {\bibinfo
  {journal} {J. Lightwave Technol.}\ }\textbf {\bibinfo {volume} {24}},\
  \bibinfo {pages} {2932} (\bibinfo {year} {2006})}\BibitemShut {NoStop}%
\bibitem [{\citenamefont {Chen}\ and\ \citenamefont
  {Murphy}(2016)}]{chen:2016}%
  \BibitemOpen
  \bibfield  {author} {\bibinfo {author} {\bibfnamefont {A.}~\bibnamefont
  {Chen}}\ and\ \bibinfo {author} {\bibfnamefont {E.}~\bibnamefont {Murphy}},\
  }\href {https://books.google.de/books?id=lTTNBQAAQBAJ} {\emph {\bibinfo
  {title} {Broadband Optical Modulators: Science, Technology, and
  Applications}}}\ (\bibinfo  {publisher} {CRC Press},\ \bibinfo {address}
  {Boca Raton},\ \bibinfo {year} {2016})\BibitemShut {NoStop}%
\bibitem [{\citenamefont {Karski}\ \emph {et~al.}(2011)\citenamefont {Karski},
  \citenamefont {F{\"o}rster}, \citenamefont {Choi}, \citenamefont {Alt},
  \citenamefont {Alberti}, \citenamefont {Widera},\ and\ \citenamefont
  {Meschede}}]{Karski:2011}%
  \BibitemOpen
  \bibfield  {author} {\bibinfo {author} {\bibfnamefont {M.}~\bibnamefont
  {Karski}}, \bibinfo {author} {\bibfnamefont {L.}~\bibnamefont {F{\"o}rster}},
  \bibinfo {author} {\bibfnamefont {J.}~\bibnamefont {Choi}}, \bibinfo {author}
  {\bibfnamefont {W.}~\bibnamefont {Alt}}, \bibinfo {author} {\bibfnamefont
  {A.}~\bibnamefont {Alberti}}, \bibinfo {author} {\bibfnamefont
  {A.}~\bibnamefont {Widera}}, \ and\ \bibinfo {author} {\bibfnamefont
  {D.}~\bibnamefont {Meschede}},\ }\bibfield  {title} {\enquote {\bibinfo
  {title} {{Direct Observation and Analysis of Spin-Dependent Transport of
  Single Atoms in a 1D Optical Lattice}},}\ }\href
  {http://dx.doi.org/10.3938/jkps.59.2947} {\bibfield  {journal} {\bibinfo
  {journal} {J. Korean Phys. Soc.}\ }\textbf {\bibinfo {volume} {59}},\
  \bibinfo {pages} {2947} (\bibinfo {year} {2011})}\BibitemShut {NoStop}%
\bibitem [{\citenamefont {Alberti}\ \emph {et~al.}(2014)\citenamefont
  {Alberti}, \citenamefont {Alt}, \citenamefont {Werner},\ and\ \citenamefont
  {Meschede}}]{Alberti14}%
  \BibitemOpen
  \bibfield  {author} {\bibinfo {author} {\bibfnamefont {A.}~\bibnamefont
  {Alberti}}, \bibinfo {author} {\bibfnamefont {W.}~\bibnamefont {Alt}},
  \bibinfo {author} {\bibfnamefont {R.}~\bibnamefont {Werner}}, \ and\ \bibinfo
  {author} {\bibfnamefont {D.}~\bibnamefont {Meschede}},\ }\bibfield  {title}
  {\enquote {\bibinfo {title} {{Decoherence Models for Discrete-Time Quantum
  Walks and their Application to Neutral Atom Experiments}},}\ }\href
  {http://dx.doi.org/10.1088/1367-2630/16/12/123052} {\bibfield  {journal}
  {\bibinfo  {journal} {New J. Phys.}\ }\textbf {\bibinfo {volume} {16}},\
  \bibinfo {pages} {123052} (\bibinfo {year} {2014})}\BibitemShut {NoStop}%
\bibitem [{\citenamefont {Robens}\ \emph
  {et~al.}(2017{\natexlab{a}})\citenamefont {Robens}, \citenamefont {Zopes},
  \citenamefont {Alt}, \citenamefont {Brakhane}, \citenamefont {Meschede},\
  and\ \citenamefont {Alberti}}]{Robens:2016bf}%
  \BibitemOpen
  \bibfield  {author} {\bibinfo {author} {\bibfnamefont {C.}~\bibnamefont
  {Robens}}, \bibinfo {author} {\bibfnamefont {J.}~\bibnamefont {Zopes}},
  \bibinfo {author} {\bibfnamefont {W.}~\bibnamefont {Alt}}, \bibinfo {author}
  {\bibfnamefont {S.}~\bibnamefont {Brakhane}}, \bibinfo {author}
  {\bibfnamefont {D.}~\bibnamefont {Meschede}}, \ and\ \bibinfo {author}
  {\bibfnamefont {A.}~\bibnamefont {Alberti}},\ }\bibfield  {title} {\enquote
  {\bibinfo {title} {{Low-Entropy States of Neutral Atoms in
  Polarization-Synthesized Optical Lattices}},}\ }\href
  {http://dx.doi.org/10.1103/PhysRevLett.118.065302} {\bibfield  {journal}
  {\bibinfo  {journal} {Phys. Rev. Lett.}\ }\textbf {\bibinfo {volume} {118}},\
  \bibinfo {pages} {065302} (\bibinfo {year} {2017}{\natexlab{a}})}\BibitemShut
  {NoStop}%
\bibitem [{\citenamefont {Robens}\ \emph {et~al.}(2015)\citenamefont {Robens},
  \citenamefont {Alt}, \citenamefont {Meschede}, \citenamefont {Emary},\ and\
  \citenamefont {Alberti}}]{Robens:2015}%
  \BibitemOpen
  \bibfield  {author} {\bibinfo {author} {\bibfnamefont {C.}~\bibnamefont
  {Robens}}, \bibinfo {author} {\bibfnamefont {W.}~\bibnamefont {Alt}},
  \bibinfo {author} {\bibfnamefont {D.}~\bibnamefont {Meschede}}, \bibinfo
  {author} {\bibfnamefont {C.}~\bibnamefont {Emary}}, \ and\ \bibinfo {author}
  {\bibfnamefont {A.}~\bibnamefont {Alberti}},\ }\bibfield  {title} {\enquote
  {\bibinfo {title} {{Ideal Negative Measurements in Quantum Walks Disprove
  Theories Based on Classical Trajectories}},}\ }\href
  {http://dx.doi.org/10.1103/PhysRevX.5.011003} {\bibfield  {journal} {\bibinfo
   {journal} {Phys. Rev. X}\ }\textbf {\bibinfo {volume} {5}},\ \bibinfo
  {pages} {011003} (\bibinfo {year} {2015})}\BibitemShut {NoStop}%
\bibitem [{\citenamefont {Robens}\ \emph
  {et~al.}(2017{\natexlab{b}})\citenamefont {Robens}, \citenamefont {Alt},
  \citenamefont {Emary}, \citenamefont {Meschede},\ and\ \citenamefont
  {Alberti}}]{Robens:2016uz}%
  \BibitemOpen
  \bibfield  {author} {\bibinfo {author} {\bibfnamefont {C.}~\bibnamefont
  {Robens}}, \bibinfo {author} {\bibfnamefont {W.}~\bibnamefont {Alt}},
  \bibinfo {author} {\bibfnamefont {C.}~\bibnamefont {Emary}}, \bibinfo
  {author} {\bibfnamefont {D.}~\bibnamefont {Meschede}}, \ and\ \bibinfo
  {author} {\bibfnamefont {A.}~\bibnamefont {Alberti}},\ }\bibfield  {title}
  {\enquote {\bibinfo {title} {{Atomic `bomb testing': the Elitzur-Vaidman
  experiment violates the Leggett-Garg inequality}},}\ }\href
  {http://dx.doi.org/10.1007/s00340-016-6581-y} {\bibfield  {journal} {\bibinfo
   {journal} {Appl. Phys. B}\ }\textbf {\bibinfo {volume} {123}},\ \bibinfo
  {pages} {12} (\bibinfo {year} {2017}{\natexlab{b}})}\BibitemShut {NoStop}%
\bibitem [{\citenamefont {Murphy}\ \emph {et~al.}(2009)\citenamefont {Murphy},
  \citenamefont {Jiang}, \citenamefont {Khaneja},\ and\ \citenamefont
  {Calarco}}]{Murphy:2009}%
  \BibitemOpen
  \bibfield  {author} {\bibinfo {author} {\bibfnamefont {M.}~\bibnamefont
  {Murphy}}, \bibinfo {author} {\bibfnamefont {L.}~\bibnamefont {Jiang}},
  \bibinfo {author} {\bibfnamefont {N.}~\bibnamefont {Khaneja}}, \ and\
  \bibinfo {author} {\bibfnamefont {T.}~\bibnamefont {Calarco}},\ }\bibfield
  {title} {\enquote {\bibinfo {title} {{High-fidelity fast quantum transport
  with imperfect controls}},}\ }\href
  {http://dx.doi.org/10.1103/PhysRevA.79.020301} {\bibfield  {journal}
  {\bibinfo  {journal} {Phys. Rev. A}\ }\textbf {\bibinfo {volume} {79}},\
  \bibinfo {pages} {020301} (\bibinfo {year} {2009})}\BibitemShut {NoStop}%
\bibitem [{\citenamefont {Torrontegui}\ \emph {et~al.}(2011)\citenamefont
  {Torrontegui}, \citenamefont {Ib{\'a}{\~n}ez}, \citenamefont {Chen},
  \citenamefont {Ruschhaupt}, \citenamefont {Gu{\'e}ry-Odelin},\ and\
  \citenamefont {Muga}}]{Torrontegui:2011}%
  \BibitemOpen
  \bibfield  {author} {\bibinfo {author} {\bibfnamefont {E.}~\bibnamefont
  {Torrontegui}}, \bibinfo {author} {\bibfnamefont {S.}~\bibnamefont
  {Ib{\'a}{\~n}ez}}, \bibinfo {author} {\bibfnamefont {X.}~\bibnamefont
  {Chen}}, \bibinfo {author} {\bibfnamefont {A.}~\bibnamefont {Ruschhaupt}},
  \bibinfo {author} {\bibfnamefont {D.}~\bibnamefont {Gu{\'e}ry-Odelin}}, \
  and\ \bibinfo {author} {\bibfnamefont {J.~G.}\ \bibnamefont {Muga}},\
  }\bibfield  {title} {\enquote {\bibinfo {title} {{Fast atomic transport
  without vibrational heating}},}\ }\href
  {http://dx.doi.org/10.1103/physreva.83.013415} {\bibfield  {journal}
  {\bibinfo  {journal} {Phys. Rev. A}\ }\textbf {\bibinfo {volume} {83}},\
  \bibinfo {pages} {013415} (\bibinfo {year} {2011})}\BibitemShut {NoStop}%
\bibitem [{\citenamefont {Roos}\ \emph {et~al.}(2017)\citenamefont {Roos},
  \citenamefont {Alberti}, \citenamefont {Meschede}, \citenamefont {Hauke},\
  and\ \citenamefont {H{\"a}ffner}}]{Roos:2017}%
  \BibitemOpen
  \bibfield  {author} {\bibinfo {author} {\bibfnamefont {C.~F.}\ \bibnamefont
  {Roos}}, \bibinfo {author} {\bibfnamefont {A.}~\bibnamefont {Alberti}},
  \bibinfo {author} {\bibfnamefont {D.}~\bibnamefont {Meschede}}, \bibinfo
  {author} {\bibfnamefont {P.}~\bibnamefont {Hauke}}, \ and\ \bibinfo {author}
  {\bibfnamefont {H.}~\bibnamefont {H{\"a}ffner}},\ }\bibfield  {title}
  {\enquote {\bibinfo {title} {{Revealing Quantum Statistics with a Pair of
  Distant Atoms}},}\ }\href {http://dx.doi.org/10.1103/PhysRevLett.119.160401}
  {\bibfield  {journal} {\bibinfo  {journal} {Phys. Rev. Lett.}\ }\textbf
  {\bibinfo {volume} {119}},\ \bibinfo {pages} {160401} (\bibinfo {year}
  {2017})}\BibitemShut {NoStop}%
\bibitem [{\citenamefont {Dorner}\ \emph {et~al.}(2013)\citenamefont {Dorner},
  \citenamefont {Clark}, \citenamefont {Heaney}, \citenamefont {Fazio},
  \citenamefont {Goold},\ and\ \citenamefont {Vedral}}]{Dorner:2013}%
  \BibitemOpen
  \bibfield  {author} {\bibinfo {author} {\bibfnamefont {R.}~\bibnamefont
  {Dorner}}, \bibinfo {author} {\bibfnamefont {S.~R.}\ \bibnamefont {Clark}},
  \bibinfo {author} {\bibfnamefont {L.}~\bibnamefont {Heaney}}, \bibinfo
  {author} {\bibfnamefont {R.}~\bibnamefont {Fazio}}, \bibinfo {author}
  {\bibfnamefont {J.}~\bibnamefont {Goold}}, \ and\ \bibinfo {author}
  {\bibfnamefont {V.}~\bibnamefont {Vedral}},\ }\bibfield  {title} {\enquote
  {\bibinfo {title} {{Extracting Quantum Work Statistics and Fluctuation
  Theorems by Single-Qubit Interferometry}},}\ }\href
  {http://dx.doi.org/10.1103/PhysRevLett.110.230601} {\bibfield  {journal}
  {\bibinfo  {journal} {Phys. Rev. Lett.}\ }\textbf {\bibinfo {volume} {110}},\
  \bibinfo {pages} {230601} (\bibinfo {year} {2013})}\BibitemShut {NoStop}%
\bibitem [{\citenamefont {Horstmann}\ \emph {et~al.}(2010)\citenamefont
  {Horstmann}, \citenamefont {D{\"u}rr},\ and\ \citenamefont
  {Roscilde}}]{Horstmann:2010}%
  \BibitemOpen
  \bibfield  {author} {\bibinfo {author} {\bibfnamefont {B.}~\bibnamefont
  {Horstmann}}, \bibinfo {author} {\bibfnamefont {S.}~\bibnamefont {D{\"u}rr}},
  \ and\ \bibinfo {author} {\bibfnamefont {T.}~\bibnamefont {Roscilde}},\
  }\bibfield  {title} {\enquote {\bibinfo {title} {{Localization of Cold Atoms
  in State-Dependent Optical Lattices via a Rabi Pulse}},}\ }\href
  {http://dx.doi.org/10.1103/PhysRevLett.105.160402} {\bibfield  {journal}
  {\bibinfo  {journal} {Phys. Rev. Lett.}\ }\textbf {\bibinfo {volume} {105}},\
  \bibinfo {pages} {160402} (\bibinfo {year} {2010})}\BibitemShut {NoStop}%
\bibitem [{\citenamefont {de~Vega}\ \emph {et~al.}(2008)\citenamefont
  {de~Vega}, \citenamefont {Porras},\ and\ \citenamefont
  {Ignacio~Cirac}}]{deVega:2008}%
  \BibitemOpen
  \bibfield  {author} {\bibinfo {author} {\bibfnamefont {I.}~\bibnamefont
  {de~Vega}}, \bibinfo {author} {\bibfnamefont {D.}~\bibnamefont {Porras}}, \
  and\ \bibinfo {author} {\bibfnamefont {J.}~\bibnamefont {Ignacio~Cirac}},\
  }\bibfield  {title} {\enquote {\bibinfo {title} {{Matter-Wave Emission in
  Optical Lattices: Single Particle and Collective Effects}},}\ }\href
  {http://dx.doi.org/10.1103/PhysRevLett.101.260404} {\bibfield  {journal}
  {\bibinfo  {journal} {Phys. Rev. Lett.}\ }\textbf {\bibinfo {volume} {101}},\
  \bibinfo {pages} {260404} (\bibinfo {year} {2008})}\BibitemShut {NoStop}%
\bibitem [{\citenamefont {Lan}\ and\ \citenamefont {Lobo}(2014)}]{Carlos:2014}%
  \BibitemOpen
  \bibfield  {author} {\bibinfo {author} {\bibfnamefont {Z.}~\bibnamefont
  {Lan}}\ and\ \bibinfo {author} {\bibfnamefont {C.}~\bibnamefont {Lobo}},\
  }\bibfield  {title} {\enquote {\bibinfo {title} {{Optical lattices with large
  scattering length: Using few-body physics to simulate an electron-phonon
  system}},}\ }\href {http://dx.doi.org/10.1103/PhysRevA.90.033627} {\bibfield
  {journal} {\bibinfo  {journal} {Phys. Rev. A}\ }\textbf {\bibinfo {volume}
  {90}},\ \bibinfo {pages} {033627} (\bibinfo {year} {2014})}\BibitemShut
  {NoStop}%
\bibitem [{\citenamefont {Shi}\ \emph {et~al.}(2016)\citenamefont {Shi},
  \citenamefont {Wu}, \citenamefont {Gonz{\'a}lez-Tudela},\ and\ \citenamefont
  {Cirac}}]{Shi:2016}%
  \BibitemOpen
  \bibfield  {author} {\bibinfo {author} {\bibfnamefont {T.}~\bibnamefont
  {Shi}}, \bibinfo {author} {\bibfnamefont {Y.-H.}\ \bibnamefont {Wu}},
  \bibinfo {author} {\bibfnamefont {A.}~\bibnamefont {Gonz{\'a}lez-Tudela}}, \
  and\ \bibinfo {author} {\bibfnamefont {J.~I.}\ \bibnamefont {Cirac}},\
  }\bibfield  {title} {\enquote {\bibinfo {title} {{Bound States in Boson
  Impurity Models}},}\ }\href {http://dx.doi.org/10.1103/PhysRevX.6.021027}
  {\bibfield  {journal} {\bibinfo  {journal} {Phys. Rev. X}\ }\textbf {\bibinfo
  {volume} {6}},\ \bibinfo {pages} {021027} (\bibinfo {year}
  {2016})}\BibitemShut {NoStop}%
\bibitem [{\citenamefont {Auzinsh}\ \emph {et~al.}(2010)\citenamefont
  {Auzinsh}, \citenamefont {Budker},\ and\ \citenamefont
  {Rochester}}]{Budker:2010}%
  \BibitemOpen
  \bibfield  {author} {\bibinfo {author} {\bibfnamefont {M.}~\bibnamefont
  {Auzinsh}}, \bibinfo {author} {\bibfnamefont {D.}~\bibnamefont {Budker}}, \
  and\ \bibinfo {author} {\bibfnamefont {S.~M.}\ \bibnamefont {Rochester}},\
  }\href@noop {} {\emph {\bibinfo {title} {{Optically Polarized Atoms:
  Understanding light-atom interactions}}}}\ (\bibinfo  {publisher} {Oxford
  University Press},\ \bibinfo {address} {New York},\ \bibinfo {year}
  {2010})\BibitemShut {NoStop}%
\bibitem [{\citenamefont {Collett}(2005)}]{Collett:2005}%
  \BibitemOpen
  \bibfield  {author} {\bibinfo {author} {\bibfnamefont {E.}~\bibnamefont
  {Collett}},\ }\href {http://dx.doi.org/10.1117/3.626141} {\emph {\bibinfo
  {title} {Field Guide to Polarization}}}\ (\bibinfo  {publisher}
  {{SPIE\textemdash{}International Society for Optical Engineering}},\ \bibinfo
  {address} {Bellingham, WA},\ \bibinfo {year} {2005})\BibitemShut {NoStop}%
\bibitem [{\citenamefont {Legr{\'e}}\ \emph {et~al.}(2003)\citenamefont
  {Legr{\'e}}, \citenamefont {Wegm{\"u}ller},\ and\ \citenamefont
  {Gisin}}]{Legre:2003}%
  \BibitemOpen
  \bibfield  {author} {\bibinfo {author} {\bibfnamefont {M.}~\bibnamefont
  {Legr{\'e}}}, \bibinfo {author} {\bibfnamefont {M.}~\bibnamefont
  {Wegm{\"u}ller}}, \ and\ \bibinfo {author} {\bibfnamefont {N.}~\bibnamefont
  {Gisin}},\ }\bibfield  {title} {\enquote {\bibinfo {title} {{Quantum
  measurement of the degree of polarization of a light beam.}}}\ }\href
  {http://dx.doi.org/10.1103/PhysRevLett.91.167902} {\bibfield  {journal}
  {\bibinfo  {journal} {Phys. Rev. Lett.}\ }\textbf {\bibinfo {volume} {91}},\
  \bibinfo {pages} {167902} (\bibinfo {year} {2003})}\BibitemShut {NoStop}%
\bibitem [{\citenamefont {Varnham}\ \emph {et~al.}(1984)\citenamefont
  {Varnham}, \citenamefont {Payne},\ and\ \citenamefont {Love}}]{Varnham:1984}%
  \BibitemOpen
  \bibfield  {author} {\bibinfo {author} {\bibfnamefont {M.~P.}\ \bibnamefont
  {Varnham}}, \bibinfo {author} {\bibfnamefont {D.~N.}\ \bibnamefont {Payne}},
  \ and\ \bibinfo {author} {\bibfnamefont {J.~D.}\ \bibnamefont {Love}},\
  }\bibfield  {title} {\enquote {\bibinfo {title} {{Fundamental limits to the
  transmission of linearly polarised light by birefringent optical fibres}},}\
  }\href {http://dx.doi.org/10.1049/el:19840038} {\bibfield  {journal}
  {\bibinfo  {journal} {Electron. Lett.}\ }\textbf {\bibinfo {volume} {20}},\
  \bibinfo {pages} {55} (\bibinfo {year} {1984})}\BibitemShut {NoStop}%
\bibitem [{\citenamefont {Takada}\ \emph {et~al.}(1986)\citenamefont {Takada},
  \citenamefont {Okamoto}, \citenamefont {Sasaki},\ and\ \citenamefont
  {Noda}}]{Takada:1986}%
  \BibitemOpen
  \bibfield  {author} {\bibinfo {author} {\bibfnamefont {K.}~\bibnamefont
  {Takada}}, \bibinfo {author} {\bibfnamefont {K.}~\bibnamefont {Okamoto}},
  \bibinfo {author} {\bibfnamefont {Y.}~\bibnamefont {Sasaki}}, \ and\ \bibinfo
  {author} {\bibfnamefont {J.}~\bibnamefont {Noda}},\ }\bibfield  {title}
  {\enquote {\bibinfo {title} {{Ultimate limit of polarization cross talk in
  birefringent polarization-maintaining fibers}},}\ }\href
  {http://dx.doi.org/10.1364/JOSAA.3.001594} {\bibfield  {journal} {\bibinfo
  {journal} {J. Opt. Soc. Am. A}\ }\textbf {\bibinfo {volume} {3}},\ \bibinfo
  {pages} {1594} (\bibinfo {year} {1986})}\BibitemShut {NoStop}%
\bibitem [{\citenamefont {Sears}(1990)}]{Sears:1990}%
  \BibitemOpen
  \bibfield  {author} {\bibinfo {author} {\bibfnamefont {F.~M.}\ \bibnamefont
  {Sears}},\ }\bibfield  {title} {\enquote {\bibinfo {title}
  {{Polarization-maintenance limits in polarization-maintaining fibers and
  measurements}},}\ }\href {http://dx.doi.org/10.1109/50.54475} {\bibfield
  {journal} {\bibinfo  {journal} {J. Lightwave Technol.}\ }\textbf {\bibinfo
  {volume} {8}},\ \bibinfo {pages} {684} (\bibinfo {year} {1990})}\BibitemShut
  {NoStop}%
\bibitem [{\citenamefont {Belmechri}\ \emph {et~al.}(2013)\citenamefont
  {Belmechri}, \citenamefont {F{\"o}rster}, \citenamefont {Alt}, \citenamefont
  {Widera}, \citenamefont {Meschede},\ and\ \citenamefont
  {Alberti}}]{Belmechri:2013}%
  \BibitemOpen
  \bibfield  {author} {\bibinfo {author} {\bibfnamefont {N.}~\bibnamefont
  {Belmechri}}, \bibinfo {author} {\bibfnamefont {L.}~\bibnamefont
  {F{\"o}rster}}, \bibinfo {author} {\bibfnamefont {W.}~\bibnamefont {Alt}},
  \bibinfo {author} {\bibfnamefont {A.}~\bibnamefont {Widera}}, \bibinfo
  {author} {\bibfnamefont {D.}~\bibnamefont {Meschede}}, \ and\ \bibinfo
  {author} {\bibfnamefont {A.}~\bibnamefont {Alberti}},\ }\bibfield  {title}
  {\enquote {\bibinfo {title} {{Microwave control of atomic motional states in
  a spin-dependent optical lattice}},}\ }\href
  {http://dx.doi.org/10.1088/0953-4075/46/10/104006} {\bibfield  {journal}
  {\bibinfo  {journal} {J. Phys. B: At. Mol. Phys.}\ }\textbf {\bibinfo
  {volume} {46}},\ \bibinfo {pages} {104006} (\bibinfo {year}
  {2013})}\BibitemShut {NoStop}%
\bibitem [{\citenamefont {Kuhr}\ \emph {et~al.}(2005)\citenamefont {Kuhr},
  \citenamefont {Alt}, \citenamefont {Schrader}, \citenamefont {Dotsenko},
  \citenamefont {Miroshnychenko}, \citenamefont {Rauschenbeutel},\ and\
  \citenamefont {Meschede}}]{Kuhr:2005}%
  \BibitemOpen
  \bibfield  {author} {\bibinfo {author} {\bibfnamefont {S.}~\bibnamefont
  {Kuhr}}, \bibinfo {author} {\bibfnamefont {W.}~\bibnamefont {Alt}}, \bibinfo
  {author} {\bibfnamefont {D.}~\bibnamefont {Schrader}}, \bibinfo {author}
  {\bibfnamefont {I.}~\bibnamefont {Dotsenko}}, \bibinfo {author}
  {\bibfnamefont {Y.}~\bibnamefont {Miroshnychenko}}, \bibinfo {author}
  {\bibfnamefont {A.}~\bibnamefont {Rauschenbeutel}}, \ and\ \bibinfo {author}
  {\bibfnamefont {D.}~\bibnamefont {Meschede}},\ }\bibfield  {title} {\enquote
  {\bibinfo {title} {{Analysis of dephasing mechanisms in a standing-wave
  dipole trap}},}\ }\href {http://dx.doi.org/10.1103/PhysRevA.72.023406}
  {\bibfield  {journal} {\bibinfo  {journal} {Phys. Rev. A}\ }\textbf {\bibinfo
  {volume} {72}},\ \bibinfo {pages} {023406} (\bibinfo {year}
  {2005})}\BibitemShut {NoStop}%
\bibitem [{\citenamefont {Gehm}\ \emph {et~al.}(1998)\citenamefont {Gehm},
  \citenamefont {O'Hara}, \citenamefont {Savard},\ and\ \citenamefont
  {Thomas}}]{Gehm:1998}%
  \BibitemOpen
  \bibfield  {author} {\bibinfo {author} {\bibfnamefont {M.~E.}\ \bibnamefont
  {Gehm}}, \bibinfo {author} {\bibfnamefont {K.~M.}\ \bibnamefont {O'Hara}},
  \bibinfo {author} {\bibfnamefont {T.~A.}\ \bibnamefont {Savard}}, \ and\
  \bibinfo {author} {\bibfnamefont {J.~E.}\ \bibnamefont {Thomas}},\ }\bibfield
   {title} {\enquote {\bibinfo {title} {{Dynamics of noise-induced heating in
  atom traps}},}\ }\href {http://dx.doi.org/10.1103/PhysRevA.58.3914}
  {\bibfield  {journal} {\bibinfo  {journal} {Phys. Rev. A}\ }\textbf {\bibinfo
  {volume} {58}},\ \bibinfo {pages} {3914} (\bibinfo {year}
  {1998})}\BibitemShut {NoStop}%
\bibitem [{\citenamefont {Gibbons}\ \emph {et~al.}(2008)\citenamefont
  {Gibbons}, \citenamefont {Kim}, \citenamefont {Fortier}, \citenamefont
  {Ahmadi},\ and\ \citenamefont {Chapman}}]{Gibbons:2008}%
  \BibitemOpen
  \bibfield  {author} {\bibinfo {author} {\bibfnamefont {M.~J.}\ \bibnamefont
  {Gibbons}}, \bibinfo {author} {\bibfnamefont {S.~Y.}\ \bibnamefont {Kim}},
  \bibinfo {author} {\bibfnamefont {K.~M.}\ \bibnamefont {Fortier}}, \bibinfo
  {author} {\bibfnamefont {P.}~\bibnamefont {Ahmadi}}, \ and\ \bibinfo {author}
  {\bibfnamefont {M.~S.}\ \bibnamefont {Chapman}},\ }\bibfield  {title}
  {\enquote {\bibinfo {title} {{Achieving very long lifetimes in optical
  lattices with pulsed cooling}},}\ }\href
  {http://dx.doi.org/10.1103/PhysRevA.78.043418} {\bibfield  {journal}
  {\bibinfo  {journal} {Phys. Rev. A}\ }\textbf {\bibinfo {volume} {78}},\
  \bibinfo {pages} {043418} (\bibinfo {year} {2008})}\BibitemShut {NoStop}%
\bibitem [{\citenamefont {Blatt}\ \emph {et~al.}(2015)\citenamefont {Blatt},
  \citenamefont {Mazurenko}, \citenamefont {Parsons}, \citenamefont {Chiu},
  \citenamefont {Huber},\ and\ \citenamefont {Greiner}}]{Blatt:2015}%
  \BibitemOpen
  \bibfield  {author} {\bibinfo {author} {\bibfnamefont {S.}~\bibnamefont
  {Blatt}}, \bibinfo {author} {\bibfnamefont {A.}~\bibnamefont {Mazurenko}},
  \bibinfo {author} {\bibfnamefont {M.~F.}\ \bibnamefont {Parsons}}, \bibinfo
  {author} {\bibfnamefont {C.~S.}\ \bibnamefont {Chiu}}, \bibinfo {author}
  {\bibfnamefont {F.}~\bibnamefont {Huber}}, \ and\ \bibinfo {author}
  {\bibfnamefont {M.}~\bibnamefont {Greiner}},\ }\bibfield  {title} {\enquote
  {\bibinfo {title} {{Low-noise optical lattices for ultracold ${}^6$Li}},}\
  }\href {http://dx.doi.org/10.1103/PhysRevA.92.021402} {\bibfield  {journal}
  {\bibinfo  {journal} {Phys. Rev. A}\ }\textbf {\bibinfo {volume} {92}},\
  \bibinfo {pages} {021402} (\bibinfo {year} {2015})}\BibitemShut {NoStop}%
\bibitem [{\citenamefont {Walther}\ \emph {et~al.}(2012)\citenamefont
  {Walther}, \citenamefont {Ziesel}, \citenamefont {Ruster}, \citenamefont
  {Dawkins}, \citenamefont {Ott}, \citenamefont {Hettrich}, \citenamefont
  {Singer}, \citenamefont {Schmidt-Kaler},\ and\ \citenamefont
  {Poschinger}}]{Walther:2012}%
  \BibitemOpen
  \bibfield  {author} {\bibinfo {author} {\bibfnamefont {A.}~\bibnamefont
  {Walther}}, \bibinfo {author} {\bibfnamefont {F.}~\bibnamefont {Ziesel}},
  \bibinfo {author} {\bibfnamefont {T.}~\bibnamefont {Ruster}}, \bibinfo
  {author} {\bibfnamefont {S.~T.}\ \bibnamefont {Dawkins}}, \bibinfo {author}
  {\bibfnamefont {K.}~\bibnamefont {Ott}}, \bibinfo {author} {\bibfnamefont
  {M.}~\bibnamefont {Hettrich}}, \bibinfo {author} {\bibfnamefont
  {K.}~\bibnamefont {Singer}}, \bibinfo {author} {\bibfnamefont
  {F.}~\bibnamefont {Schmidt-Kaler}}, \ and\ \bibinfo {author} {\bibfnamefont
  {U.}~\bibnamefont {Poschinger}},\ }\bibfield  {title} {\enquote {\bibinfo
  {title} {{Controlling Fast Transport of Cold Trapped Ions}},}\ }\href
  {http://dx.doi.org/10.1103/PhysRevLett.109.080501} {\bibfield  {journal}
  {\bibinfo  {journal} {Phys. Rev. Lett.}\ }\textbf {\bibinfo {volume} {109}},\
  \bibinfo {pages} {080501} (\bibinfo {year} {2012})}\BibitemShut {NoStop}%
\bibitem [{Ext()}]{ExtinctionRatio}%
  \BibitemOpen
  \bibinfo {note} {We define the extinction ratio as $\eta =
  I_\text{min}/I_\text{max}$, where $I_\text{max}$ ($I_\text{min}$) is the
  maximum (minimum) laser intensity transmitted through a rotating linear
  polarizer.}\BibitemShut {Stop}%
\bibitem [{DOP()}]{DOP}%
  \BibitemOpen
  \bibinfo {note} {We denote the density matrix describing the polarization
  state after the two plates by \encapsulateMath{$\rho =
  (\sigma_0+\vec\sigma\cdot\vec{S}/S_0)$}, where the direction of
  \encapsulateMath{$\vec{S}$} is arbitrarily adjustable through the two plates,
  \encapsulateMath{$\vec{\sigma}$} is the vector of Pauli matrices, and
  $\sigma_0$ is the identity matrix. By projecting on H polarization, we obtain
  that the minimum of $\eta = \tr[\rho (\sigma_0+\sigma_1)/2]=(1+S_1/S_0)/2$
  occurs when \encapsulateMath{$\vec{S}=S_0(-\DOP,0,0)$}, from which
  Eq.~(\ref{eqn:etaDOP}) directly follows.}\BibitemShut {Stop}%
\bibitem [{Sin()}]{SingleSidebandPhaseNoise}%
  \BibitemOpen
  \bibinfo {note} {We follow the convention to represent the phase noise
  spectral density, $S(\nu)$, as single-sideband phase noise, $\mathcal{L}(\nu)
  = 10\log_{10}[S(\nu)/2\;\si{\hertz}/\si{\radian^2}]$, expressed in units of
  \si{\dBc/\hertz}.}\BibitemShut {Stop}%
\bibitem [{\citenamefont {Stenholm}(1986)}]{Stenholm:1986}%
  \BibitemOpen
  \bibfield  {author} {\bibinfo {author} {\bibfnamefont {S.}~\bibnamefont
  {Stenholm}},\ }\bibfield  {title} {\enquote {\bibinfo {title} {{The
  semiclassical theory of laser cooling}},}\ }\href
  {http://dx.doi.org/10.1103/RevModPhys.58.699} {\bibfield  {journal} {\bibinfo
   {journal} {Rev. Mod. Phys.}\ }\textbf {\bibinfo {volume} {58}},\ \bibinfo
  {pages} {699} (\bibinfo {year} {1986})}\BibitemShut {NoStop}%
\bibitem [{\citenamefont {Alt}\ \emph {et~al.}(2003)\citenamefont {Alt},
  \citenamefont {Schrader}, \citenamefont {Kuhr}, \citenamefont {M{\"u}ller},
  \citenamefont {Gomer},\ and\ \citenamefont {Meschede}}]{Alt:2003}%
  \BibitemOpen
  \bibfield  {author} {\bibinfo {author} {\bibfnamefont {W.}~\bibnamefont
  {Alt}}, \bibinfo {author} {\bibfnamefont {D.}~\bibnamefont {Schrader}},
  \bibinfo {author} {\bibfnamefont {S.}~\bibnamefont {Kuhr}}, \bibinfo {author}
  {\bibfnamefont {M.}~\bibnamefont {M{\"u}ller}}, \bibinfo {author}
  {\bibfnamefont {V.}~\bibnamefont {Gomer}}, \ and\ \bibinfo {author}
  {\bibfnamefont {D.}~\bibnamefont {Meschede}},\ }\bibfield  {title} {\enquote
  {\bibinfo {title} {{Single atoms in a standing-wave dipole trap}},}\ }\href
  {http://dx.doi.org/10.1103/PhysRevA.67.033403} {\bibfield  {journal}
  {\bibinfo  {journal} {Phys. Rev. A}\ }\textbf {\bibinfo {volume} {67}},\
  \bibinfo {pages} {033403} (\bibinfo {year} {2003})}\BibitemShut {NoStop}%
\bibitem [{\citenamefont {F{\"o}rster}\ \emph {et~al.}(2009)\citenamefont
  {F{\"o}rster}, \citenamefont {Karski}, \citenamefont {Choi}, \citenamefont
  {Steffen}, \citenamefont {Alt}, \citenamefont {Meschede}, \citenamefont
  {Widera}, \citenamefont {Montano}, \citenamefont {Lee}, \citenamefont
  {Rakreungdet},\ and\ \citenamefont {Jessen}}]{Forster:2009}%
  \BibitemOpen
  \bibfield  {author} {\bibinfo {author} {\bibfnamefont {L.}~\bibnamefont
  {F{\"o}rster}}, \bibinfo {author} {\bibfnamefont {M.}~\bibnamefont {Karski}},
  \bibinfo {author} {\bibfnamefont {J.-M.}\ \bibnamefont {Choi}}, \bibinfo
  {author} {\bibfnamefont {A.}~\bibnamefont {Steffen}}, \bibinfo {author}
  {\bibfnamefont {W.}~\bibnamefont {Alt}}, \bibinfo {author} {\bibfnamefont
  {D.}~\bibnamefont {Meschede}}, \bibinfo {author} {\bibfnamefont
  {A.}~\bibnamefont {Widera}}, \bibinfo {author} {\bibfnamefont
  {E.}~\bibnamefont {Montano}}, \bibinfo {author} {\bibfnamefont {J.~H.}\
  \bibnamefont {Lee}}, \bibinfo {author} {\bibfnamefont {W.}~\bibnamefont
  {Rakreungdet}}, \ and\ \bibinfo {author} {\bibfnamefont {P.~S.}\ \bibnamefont
  {Jessen}},\ }\bibfield  {title} {\enquote {\bibinfo {title} {{Microwave
  Control of Atomic Motion in Optical Lattices}},}\ }\href
  {http://dx.doi.org/10.1103/PhysRevLett.103.233001} {\bibfield  {journal}
  {\bibinfo  {journal} {Phys. Rev. Lett.}\ }\textbf {\bibinfo {volume} {103}},\
  \bibinfo {pages} {233001} (\bibinfo {year} {2009})}\BibitemShut {NoStop}%
\bibitem [{\citenamefont {Han}\ \emph {et~al.}(2000)\citenamefont {Han},
  \citenamefont {Wolf}, \citenamefont {Oliver}, \citenamefont {McCormick},
  \citenamefont {DePue},\ and\ \citenamefont {Weiss}}]{Han:2000}%
  \BibitemOpen
  \bibfield  {author} {\bibinfo {author} {\bibfnamefont {D.~J.}\ \bibnamefont
  {Han}}, \bibinfo {author} {\bibfnamefont {S.}~\bibnamefont {Wolf}}, \bibinfo
  {author} {\bibfnamefont {S.}~\bibnamefont {Oliver}}, \bibinfo {author}
  {\bibfnamefont {C.}~\bibnamefont {McCormick}}, \bibinfo {author}
  {\bibfnamefont {M.~T.}\ \bibnamefont {DePue}}, \ and\ \bibinfo {author}
  {\bibfnamefont {D.~S.}\ \bibnamefont {Weiss}},\ }\bibfield  {title} {\enquote
  {\bibinfo {title} {{3D Raman sideband cooling of cesium atoms at high
  density}},}\ }\href {http://dx.doi.org/10.1103/PhysRevLett.85.724} {\bibfield
   {journal} {\bibinfo  {journal} {Phys. Rev. Lett.}\ }\textbf {\bibinfo
  {volume} {85}},\ \bibinfo {pages} {724} (\bibinfo {year} {2000})}\BibitemShut
  {NoStop}%
\bibitem [{\citenamefont {Kaufman}\ \emph {et~al.}(2012)\citenamefont
  {Kaufman}, \citenamefont {Lester},\ and\ \citenamefont
  {Regal}}]{Kaufman:2012}%
  \BibitemOpen
  \bibfield  {author} {\bibinfo {author} {\bibfnamefont {A.~M.}\ \bibnamefont
  {Kaufman}}, \bibinfo {author} {\bibfnamefont {B.~J.}\ \bibnamefont {Lester}},
  \ and\ \bibinfo {author} {\bibfnamefont {C.~A.}\ \bibnamefont {Regal}},\
  }\bibfield  {title} {\enquote {\bibinfo {title} {{Cooling a Single Atom in an
  Optical Tweezer to Its Quantum Ground State}},}\ }\href
  {http://dx.doi.org/10.1103/PhysRevX.2.041014} {\bibfield  {journal} {\bibinfo
   {journal} {Phys. Rev. X}\ }\textbf {\bibinfo {volume} {2}},\ \bibinfo
  {pages} {041014} (\bibinfo {year} {2012})}\BibitemShut {NoStop}%
\bibitem [{\citenamefont {Thompson}\ \emph {et~al.}(2013)\citenamefont
  {Thompson}, \citenamefont {Tiecke}, \citenamefont {Zibrov}, \citenamefont
  {Vuleti{\'c}},\ and\ \citenamefont {Lukin}}]{Thompson:2013}%
  \BibitemOpen
  \bibfield  {author} {\bibinfo {author} {\bibfnamefont {J.~D.}\ \bibnamefont
  {Thompson}}, \bibinfo {author} {\bibfnamefont {T.~G.}\ \bibnamefont
  {Tiecke}}, \bibinfo {author} {\bibfnamefont {A.~S.}\ \bibnamefont {Zibrov}},
  \bibinfo {author} {\bibfnamefont {V.}~\bibnamefont {Vuleti{\'c}}}, \ and\
  \bibinfo {author} {\bibfnamefont {M.~D.}\ \bibnamefont {Lukin}},\ }\bibfield
  {title} {\enquote {\bibinfo {title} {{Coherence and Raman Sideband Cooling of
  a Single Atom in an Optical Tweezer}},}\ }\href
  {http://dx.doi.org/10.1103/PhysRevLett.110.133001} {\bibfield  {journal}
  {\bibinfo  {journal} {Phys. Rev. Lett.}\ }\textbf {\bibinfo {volume} {110}},\
  \bibinfo {pages} {133001} (\bibinfo {year} {2013})}\BibitemShut {NoStop}%
\bibitem [{\citenamefont {Leibfried}\ \emph {et~al.}(2003)\citenamefont
  {Leibfried}, \citenamefont {Blatt}, \citenamefont {Monroe},\ and\
  \citenamefont {Wineland}}]{Leibfried:2003}%
  \BibitemOpen
  \bibfield  {author} {\bibinfo {author} {\bibfnamefont {D.}~\bibnamefont
  {Leibfried}}, \bibinfo {author} {\bibfnamefont {R.}~\bibnamefont {Blatt}},
  \bibinfo {author} {\bibfnamefont {C.}~\bibnamefont {Monroe}}, \ and\ \bibinfo
  {author} {\bibfnamefont {D.}~\bibnamefont {Wineland}},\ }\bibfield  {title}
  {\enquote {\bibinfo {title} {{Quantum dynamics of single trapped ions}},}\
  }\href {http://dx.doi.org/10.1103/RevModPhys.75.281} {\bibfield  {journal}
  {\bibinfo  {journal} {Rev. Mod. Phys.}\ }\textbf {\bibinfo {volume} {75}},\
  \bibinfo {pages} {281} (\bibinfo {year} {2003})}\BibitemShut {NoStop}%
\bibitem [{Bol()}]{BoltzmannDist}%
  \BibitemOpen
  \bibinfo {note} {In the harmonic approximation, imperfect transport
  operations and lattice fluctuations excite the motional ground state to a
  coherent state. If the average number of excitations is small, the Poissonian
  distribution of the coherent state approximates an exponential Boltzmann
  distribution.}\BibitemShut {Stop}%
\end{thebibliography}%

\end{document}